

\input harvmac.tex



\input amssym.def \def\Z{{\Bbb Z}}  \def\Q{{\Bbb Q}}
\def\C{{\Bbb C}}  \def\N{{\Bbb N}}
added this macro

\def\NN{{\Bbb N}}

\def\frac#1#2{{\textstyle{#1\over #2}}}

\def\b#1{\kern-0.25pt\vbox{\hrule height 0.2pt\hbox{\vrule
width 0.2pt \kern2pt\vbox{\kern2pt \hbox{#1}\kern2pt}\kern2pt\vrule
width 0.2pt}\hrule height 0.2pt}}
\def\ST#1{\matrix{\vbox{#1}}}
\def\STrow#1{\hbox{#1}\kern-1.35pt}
\def\bv{\b{\phantom{1}}}





\def\eqalignD#1{
\vcenter{\openup1\jot\halign{
\hfil$\displaystyle{##}$~&
$\displaystyle{##}$\hfil~&
$\displaystyle{##}$\hfil\cr
#1}}
}
\def\eqalignT#1{
\vcenter{\openup1\jot\halign{
\hfil$\displaystyle{##}$~&
$\displaystyle{##}$\hfil~&
$\displaystyle{##}$\hfil~&
$\displaystyle{##}$\hfil\cr
#1}}
}

\def\eqalignQ#1{
\vcenter{\openup1\jot\halign{
\hfil$\displaystyle{##}$~&
$\displaystyle{##}$\hfil~&
$\displaystyle{##}$\hfil~&
$\displaystyle{##}$\hfil~&
$\displaystyle{##}$\hfil\cr
#1}}
}

\def\eqalignS#1{
\vcenter{\openup1\jot\halign{
\hfil$\displaystyle{##}$~&
$\displaystyle{##}$\hfil~&
$\displaystyle{##}$\hfil~&
$\displaystyle{##}$\hfil~&
$\displaystyle{##}$\hfil~&
$\displaystyle{##}$\hfil~&
$\displaystyle{##}$\hfil\cr
#1}}
}

\def\text#1{\quad\hbox{#1}\quad}
\def\gh{\hat{g}}

\def\la{\lambda}
\def\a{\alpha}
\def\e{\epsilon}
\def\nuh{{\hat \nu}}
\def\muh{{\hat \mu}}

\def\E{\widehat {E}}
\def\Ec{{\cal E}}\def\Ac{{\cal A}}
\def\A{\widehat {A}}
\def\B{\widehat {B}}
\def\C{\widehat {C}}
\def\D{\widehat {D}}

\def\F{\widehat {F}}
\def\W{\widehat {W}}
\def\rh{{\hat \rho}}
\def\lah{{\hat \lambda}}
\def\sih{{\hat \sigma}}

\def\y{{\infty}}

\def\gh{{\widehat g}}
\def\rw{\rightarrow}

\def\Om{\mathop{\Omega}\limits}

\def\Oz{{\displaystyle\Om_{=}^z}}
\def\Oxx{{\displaystyle\Om_{=}^x}}
\def\OR{{\displaystyle\Om_{=}^R}}

\def\su{\widehat{su}}

\def\sp{\widehat{sp}}

\def\max{{\rm max}}

\def\Nc{{\cal N}}

\def\Eb{\overline{E}}
\def\Ab{\overline{A}}
\def\Bb{\overline{B}}
\def\Cb{\overline{C}}
\def\Db{\overline{D}}

\def\Fb{\overline{F}}

\overfullrule=0pt

\newcount\eqnum  
\eqnum=0
\def\eq{\eqno(\secsym\the\meqno)\global\advance\meqno by1}
\def\eqlabel#1{{\xdef#1{\secsym\the\meqno}}\eq }  

\newwrite\refs 
\def\startreferences{
 \immediate\openout\refs=references
 \immediate\write\refs{\baselineskip=14pt \parindent=16pt \parskip=2pt}
}
\startreferences

\refno=0
\def\aref#1{\global\advance\refno by1
 \immediate\write\refs{\noexpand\item{\the\refno.}#1\hfil\par}}
\def\ref#1{\aref{#1}\the\refno}
\def\refname#1{\xdef#1{\the\refno}}

\newcount\exno
\exno=0
\def\Ex{\global\advance\exno by1{\noindent\sl Example \the\exno:

\nobreak\par\nobreak}}

\parskip=6pt


\Title{\vbox{\baselineskip12pt
\hbox{LAVAL-PHY-99-20}}}
{\vbox {\centerline{Generating-function method for fusion rules  }}}
\centerline{L. B\'egin$^\natural$\foot{Work supported by NSERC
(Canada).}, C. Cummins$^{\sharp 2}$ and P. Mathieu$^\natural$\foot{Work supported
by NSERC (Canada) and FCAR (Qu\'ebec).} }
\smallskip\centerline{$^\natural$ \it D\'epartement de Physique,
Universit\'e Laval, Qu\'ebec, Canada G1K 7P4}
\smallskip\centerline{$^\sharp$ \it Mathematics Department, University of
Concordia, Montr\'eal Qu\'ebec Canada H3G 1M8}

\noindent

{\bf Abstract}: This is the second of two articles devoted to
an exposition of the generating-function method for computing fusion
rules in affine Lie algebras.  The present paper focuses on fusion rules, using
the machinery developed for tensor products in the companion article. 
Although the Kac-Walton algorithm provides a
method for  constructing  a fusion generating function from the
corresponding tensor-product generating function,
we describe a  more powerful approach  which starts by
first defining the set of fusion elementary couplings from a natural extension of
the set of tensor-product elementary couplings.
A set of inequalities involving the level are derived from
this set using Farkas' lemma.  These
inequalities, taken in conjunction with the inequalities defining the tensor
products, define what we call the
{\it fusion basis}.  
Given this basis,
the  machinery of our previous paper may be  applied to construct the fusion generating
function. New generating functions for $\sp(4)$ and
$\su(4)$, together with  a closed form expression for their threshold
levels are presented.

 \Date{05/99, revised 04/00 \ \ (arXiv:hepth/0005002)}

\let\n\noindent


\newsec{Introduction}

The basic definition of a fusion coefficient is that it gives the number of
independent couplings between three different fields in conformal 
field theory (cf.
also the introduction of [\ref{L.
B\'egin, C. Cummins and P. Mathieu, {\it Generating functions for tensor products
}, hep-th/9811113 .}\refname\BCM]; for a review of conformal field theory, and
in particular fusion rules, see [\ref{P. Di Francesco, P. Mathieu, D.
S\'en\'echal, {\it Conformal Field Theory}, Springer Verlag 1997.}]). Even in
theories with a Lie group symmetry, the so-called Wess-Zumino-Witten (WZW)
models, an intrinsic conformal-field theoretical characterisation is
unavoidable.  This is  manifest in formulae for the fusion coefficients: the most
fundamental one is the Verlinde formula [\ref{E. Verlinde, Nucl. Phys. {\bf B
300} (1988) 389.}], that expresses a fusion coefficient in terms of modular $S$
matrix elements:
$${\Nc_{\lah\muh}^{(k)}}~^{\nuh} = \sum_{\sih\in
P_+^k}{S_{\lah\sih}S_{\muh\sih}S_{\nuh\sih}^*\over S_{0\sih}}\eq$$
Here we use  notation appropriate to a WZW model in
 which primary fields are in one-to-one correspondence with
the integrable representations of the spectrum-generating affine algebra at a fixed
level
$k$ (this set is denoted by
$P_+^k$) and $0$ stands for the basic representation, whose finite projection is the
scalar representation.  Fields are not distinguished from their representation
labels.
The matrix $S$ specifies
the  linear modular transformation properties of the characters of the primary fields
among themselves. Up to a constant fixed by unitarity, it takes the form
$$S_{\lah\muh}\sim \sum_{w\in W}\epsilon(w) \exp\left(-{2\pi i\over
k+g}(w(\la+\rho),
\mu+\rho)\right)\eq$$
where $g$ stands for the dual Coxeter number of the algebra under consideration,
$\rho$ is the Weyl vector, $\la$ is the finite projection of the affine weight
$\lah$  and $W$ is the finite Weyl group.

The remarkable fact that the ratio of two $S$ matrix elements is a finite
character evaluated at a special point yields a close relation between fusion
and tensor-product coefficients.  Indeed, since the finite character and its
evaluation read 
$$\chi_\la= {\sum_{w\in W} \epsilon(w) e^{w(\la+\rho)}\over\sum_{w\in W} \epsilon(w)
e^{w\rho} }\qquad
\hbox{and}\qquad \chi_\la(\xi)={\sum_{w\in W}
\epsilon(w) e^{(w(\la+\rho),\xi)}\over \sum_{w\in W} \epsilon(w)
e^{(w\rho,\xi)}}\eq$$ we observe that 
$$\chi_\la(\xi)= {S_{\lah,\sih}\over S_{0,\sih}}  \qquad {\rm with}\qquad  \xi =
-{2\pi i\over k+g}(\sigma+\rho)\eq$$ 
This leads to the Kac-Walton formula which relates
the fusion and the tensor-product coefficients.

The Verlinde formula  does not make manifest the basic integrality
property of the fusion coefficients.  The $S$ matrix elements being in general
complex numbers, it is not even clear at first sight that the fusion coefficients
are real (this follows from the unitarity property of
$S$). The integrality is ensured by the Kac-Walton formula, but in 
this case the
positivity is not manifest.

It is mainly with the aim of displaying manifestly non-negative formulae for fusion
rules that we have looked for fusion generating functions [\ref{C.J.
Cummins, P. Mathieu and M.A. Walton, Phys. Lett. {\bf B254} (1991)
390.}\refname\CMW].
 Although the construction of explicit generating functions has
an intrinsic interest, we regard 
the unravelling of the concept of {\it threshold level} - reviewed below - as
being the most important outcome of this analysis. It leads to a complete
characterisation of fusion coefficients in terms of the corresponding tensor-product
coefficients and a set of threshold levels.

As a result, the interest has shifted from the construction of fusion generating
functions to the search for threshold-level computing techniques.
For $\su(N)$, $N=2,3,4$, it has been found that the threshold level is coded
in a simple way in the Berenstein-Zelevinsky triangles [\ref{A.D.
Berenstein and A.Z. Zelevinsky, J. Algebraic Combinat. 1 (1992) 7.}\refname\BZ]
(cf. also section 7.1 of [\BCM])  describing the various distinct couplings of a
tensor product [\ref{A.N. Kirillov, P. Mathieu, D. S\'en\'echal and M. Walton,
Nucl. Phys. {\bf B391} (1993) 651.}\refname\KMSW, \ref{L. B\'egin,
A.N. Kirillov, P. Mathieu and M. Walton, Lett. Math. Phys. {\bf 28} (1993)
257.}\refname\BKMW]. However, these formulae are  difficult to
generalise to larger values of $N$.  Moreover, this approach, based on a diagrammatic
description of the tensor product, is limited to the
$\su(N)$ algebras.

The aim of the present paper is to apply 
the machinery developed in [\BCM] to these problems.
We find new generating functions for 
$\sp(4)$ and
$\su(4)$, together with  a closed form expression for their threshold
levels.  
More importantly, we introduce the  concept of
{\it fusion basis}, that is, the set of linear and  homogeneous Diophantine
inequalities that describes completely the fusion rules.


 The article is organised as follows. In section 2, after introducing
some notation, we present a brief review of
fusion rules and show, with the example of $\su(2)$, how tensor-product
generating functions and the Kac-Walton algorithm can be used to construct
fusion-rule generating functions.  A more powerful approach to the
problem is then elaborated in section 3.  It relies on the conjectural existence
of a linear and homogeneous set of inequalities that provides a complete
description of fusion rules.   Given a
set of fusion elementary couplings,  
Farkas' lemma is then  used as a technique to extract  the underlying
inequalities.  This is what we call a {\it fusion basis}, i.e., the
basis in terms of which these fusion elementary couplings are the elementary
solutions.  A complete analysis of the
$\su(3), \,\sp(4)$ and $\su(4)$  cases is presented in section 4, 5 and 6
respectively. In all three cases, the general 
expression for the
 threshold levels is obtained explicitly.
Various arguments (based on Giambelli-type formula and level-rank
duality) supporting our results are presented in Appendix A. 
In Appendix B, we recall previous conjectures and clarify their 
 relation to those
 formulated here. 


\newsec{Fusion rules}

Let $\gh$ be the  affine Lie algebra corresponding to
the finite Lie algebra $g$. Quantities with hats generally
refers to
$\gh$.  The fundamental weights of $\gh$ are denoted by
${\widehat{\omega}}_{i}$, $i=0,1,..., r$, where $r$ is the rank of $g$. 
An affine weight may be written as $$\lah=\sum_{i=0} ^{r} \lambda_{i}
{\widehat{\omega}}_{i}=[\lambda_0,\lambda_1,..., \lambda_{r}]\eq$$ If the Dynkin
labels $\la_i$ are nonnegative, the weight $\lah$  is the highest
weight of an integrable  representation of $\widehat{g}$ at level $k$, with  
 $k$  defined by
$$k=\sum_{i=0}^{r} \lambda_i a_i^{\vee}\eq$$ 
The $a_i^{\vee}$ are the co-marks: $a_0^{\vee}=1$, and the remaining
$a_i^{\vee }$ are the coefficients of expansion of the longest root of $g$
 in terms of the simple coroots.  The set of such weights
is denoted $P_+^{k}$.

To the affine
 weight $\lah$, we associate a  weight ${\lambda}$ of the finite algebra $g$
$${\lambda}=\sum_{i=1}^{r} \lambda_i
{\omega}_{i} = (\lambda_1,...,\lambda_r)\eq$$ where ${\omega}_{i}$ for $
(i=1,...,r)$ are the fundamental weights of $g$.  $\lah$ is thus uniquely fixed
from $\la$ and $k$.  The set of integrable finite weights is written $P_+$.

In the conformal field-theory context, fusion rules yield the number of
independent couplings between three given primary fields.  
Here we are
interested in fusion rules in 
WZW models [\ref{V.G.Knizhnik and A.B
Zamolodchikov, Nucl. Phys. {\bf B247} (1984) 83.},\ref{D.Gepner and E.Witten,
Nucl. Phys. {\bf B278} (1986) 493.}\refname\GW],
whose generating spectrum algebra is an
affine Lie algebra at integer level. 
 
Denote
the multiplicity of the representation $\nuh$ in the fusion rule $\lah\times\muh$
by 
$$\lah\times \muh = \sum_{\nuh\in P_+^{k}} {\Nc_{\lah\muh}^{(k)}}~^{\nuh}
\; \nuh\eq$$
and denote by ${\Nc_{\lambda\mu}}^{\nu}$ the multiplicity of the representation
$\nu$ in the tensor product $\la\otimes\mu$:
$$\la\otimes \mu = \sum_{\nu\in P_+} {\Nc_{\lambda\mu}}^{\nu}\;  \nu\eq$$
where by abuse of notation, we use the same symbol for the highest weight and
the  highest-weight representation.
The precise relation between tensor-product and fusion-rule coefficients is given
by  the Kac-Walton formula [\ref{M.A.Walton, Nucl. Phys. {\bf B340}
(1990) 777; Phys. Lett. {\bf B241} (1990) 365.}\refname\wal,\ref{V.G.Kac, {\it
Infinite dimensional Lie algebras}, 3rd edition, Cambridge Univ. Press. (1990),
exercise 13.35.},\ref{ P.Furlan, A.Ganchev and V.B. Petkova, Nucl. Phys. {\bf B343}
(1990) 205; J.Fuchs and P.van Driel, Nucl.Phys. {\bf B346} (1990) 632.}]:
$${\Nc_{\lah\muh}^{(k)}}^{\, \nuh}=
\sum_{{\xi\in P_+
\atop {w\in \W \,,\;\; w\cdot{\hat\xi}=\nuh\in P_+^{k}
}}}~{\Nc_{\lambda\mu}}^{\xi} {}~\epsilon(w) \eqlabel\kacwal$$ $w$ is an element of
the affine Weyl group $\W$, of sign $\epsilon(w)$, and the dot indicates the shifted
action,
$$w\cdot\lah=w(\lah+\rh)-\rh\qquad \quad \rh=\sum_{i=0}^r
{\widehat\omega}_{i}\eq $$ 


The Kac-Walton formula can be transformed into a simple algorithm: one first
calculates the tensor product of the corresponding finite weights and then
extends every weight to its affine version at the appropriate value of $k$ and
shift-reflects back to the integrable affine sector those weights which have
negative zeroth Dynkin label. Weights that cannot be shift-reflected in the
integrable sector are ignored  
(for example this is  the case for those which have zeroth Dynkin label
equal to
$-1$).  

The affine extension of the weights
that occur in the tensor product may not be integrable at level $k$ but  are
integrable at level $2k$.    If we divide the weight space into domains that are mapped
into each other by the application of the affine Weyl reflections, then the
 affine reflections which contribute to the Kac-Walton algorithm, apart from the
identity, are those corresponding to the domains next to 
 the fundamental
alcove and which lies in the $P_+$ cone. This is a crucial property of the
Kac-Walton algorithm for its application to the construction of fusion-rule
generating functions. Let us denote by $\W_f$ this finite subset of the affine Weyl
group that need to be considered .   For instance, the elements of
$\W_f$ for the lowest rank algebras are:
$$\eqalignD{
& \su(2): \qquad &\W_f=\{id, s_0\}  \cr
& \su(3): \qquad &\W_f=\{ id, s_0, s_1s_0, s_2s_0\}  \cr
& \su(4): \qquad & \W_f=\{id, s_0, s_1s_0,
s_3s_0,s_2s_1s_0,s_2s_3s_0,s_1s_3s_0,s_0s_1s_3s_0 \}
\cr 
& \sp(4): \qquad & \W_f=\{id, s_0, s_1s_0, s_0s_1s_0 \} \cr
& {\widehat G}_2: \qquad & \W_f=\{id, s_0, s_1s_0, s_2s_0, s_0s_2s_1s_0 \}
\cr}\eqlabel\wfund$$where
$s_i$ denotes the reflection with respect to the root $\a_i$.  This set of elements
$w$ can be characterised as follows: these are the  elements $w$ of the affine
Weyl group that satisfy the requirement:
$$w\{2\alpha_0^\vee+\alpha_1^\vee+\cdots +\alpha_r^\vee, \alpha_1^\vee, \cdots,
\alpha_r^\vee\}\in \Delta^\vee_+\eq$$
where $\Delta^\vee_+$ stands for the set of positive real coroots of the affine
algebra under consideration and
$r$ stands for its rank. This condition is adapted
from [\ref{V. Kac and M. Wakimoto, {\sl Adv. Ser. Math. Phys.} {\bf 7} (World
Scientific, 1988) 138}] as further analysed in [\ref{P. Mathieu and M.A. Walton, 
Nucl. Phys. {\bf B 553} (1999) 533-558.}].  

 Note also that (\kacwal) may be rewritten as:
  $$ \nuh \in P_+^{k}: \qquad {\Nc_{\lah\muh}^{(k)}}^{\, \nuh}=
  \sum_{w\in \W_f^{-1} \,,\;\; w\cdot \nuh\in P_+}~{\Nc_{\lambda\mu}}^{w\cdot
 \nuh} {}~\epsilon(w) \eq$$ 
 where it is understood that $w\cdot \nuh$ stands for its finite part since it
 is an index of the tensor-product coefficient. 
 This allows us to study in isolation the contribution of a single weight
 in the fusion.
 For instance, for $\su(2)$ that reads
$${\Nc_{\lah\muh}^{(k)}}^{\, \nuh} =  {\Nc_{\la\mu}}^{\, \nu}- 
{\Nc_{\la\mu}}^{\, s_0\cdot\nuh}= {\Nc_{\la_1\mu_1}}^{\, \nu_1}-
{\Nc_{\la_1\mu_1}}^{\, 2k+2-\nu_1}
\eq$$

Here is an illustrative example of the Kac-Walton algorithm that will also serve to
introduce the key notion of threshold level. Take the following $sp(4)$ tensor
product:
$(1,1)\otimes(1,1)$.  Its decomposition reads
$$(1,1)\otimes(1,1)=(0,0)\,\oplus\,(0,1)\,\oplus\,2\,(2,0)
\,\oplus\,(0,2)\,\oplus\,(0,3)\,\oplus\,2(2,1)\,\oplus\,(2,2)\,\oplus\,(4,0)\eq$$ 
The $sp(4)$ comarks are all equal to one so that the affine extension of a weight
$(m,n)$ at level $k$ is $[k-m-n,m,n]$. At level 2, the weights $(0,3)$ and
$(2,1)$ are ignored (they have $\nu_0=-1)$ and the remaining non-integrable weights
are
$[-2,2,2]$ and $[-2,4,0]$. 
Since the zeroth simple root is $\hat{\alpha}_0= [2,-2,0]$, we have
$s_0\cdot [-2,2,2]= [0,0,2]$ and $ s_0\cdot [-2,4,0] = [0,2,0]$, so that the
resulting fusion is
$$[0,1,1]\times[0,1,1]=[2,0,0]\,\oplus\,[1,0,1]\,\oplus\,[0,2,0] \eqlabel\ess$$

In the above example, we see that 
the weights $(0,0), (0,1),( 2,0)$ appear first at level 2.  It is easily checked
that they  reappear at every level $k\geq 2$.  We then say that their {\it threshold
level}, usually denoted by $k_0$, is 
$2$. 
The threshold level is thus the smallest
value of $k$ such that the fusion coefficient ${\Nc_{\lah\muh}^{(k)}}~^{\nuh}$ is
non-zero. If we indicate the threshold level by a subindex, by considering the
extension of the above tensor product at different levels, we find

$$\eqalign{ (1,1)\otimes(1,1)
=(0,0)_2 &\,\oplus\,(0,1)_2\,\oplus\,(2,0)_2\,\oplus\,(2,0)_3
\,\oplus\,(0,2)_3\cr & 
\,\oplus\,(0,3)_3\,\oplus\,2(2,1)_3\,\oplus\,(2,2)_4\,\oplus\,(4,0)_4\cr}
\eq$$ To read off a fusion at fixed level $k$, we only keep terms with index not
greater than $k$.  The concept of threshold level was first introduced in 
[\CMW]. 
 It can be shown  (cf. ref. [\KMSW]) that the existence of a threshold level
is  a consequence the depth rule of Gepner and Witten
[\GW]. The notion of threshold level
implies directly  that
$${\Nc_{\lah\muh}^{(k)}}~^{\nuh} \leq
{\Nc_{\lah\muh}^{(k+1)}}~^{\nuh}\quad \hbox{\rm and} \quad \lim_{k \rightarrow \y}
{\Nc_{\lah\muh}^{(k)}}~^{\nuh}=  {\Nc_{\lambda\mu}}^{\nu}.
\eqlabel\limformula$$

 To the 
triplet
$(\lambda,\mu,\nu)$ there corresponds
${\Nc_{\lambda\mu}}^{\nu}$ distinct couplings, hence
${\Nc_{\lambda\mu}}^{\nu}$ values of $k_{0}$, one for each distinct coupling. Let us
denote these by
$k_{0}^{(i)}, i=1, ..., {\Nc_{\lambda\mu}}^{\nu}$, implementing in this  notation
the natural ordering $k_{0}^{(i)}\leq k_{0}^{(i+1)}$. Then 
$$ {\Nc_{\lah\muh}^{(k)}}~^{\nuh} = \left\{
\eqalign{\max&(i) \text{if} k \geq {k_{0}}^{(i)} \text{and}
{\Nc_{\lambda\mu}}^{\nu}\neq 0 \cr    0 & \quad~ \text{if}
k<{k_{0}}^{(1)}
\text{or} {\Nc_{\lambda\mu}}^{\nu}=0. \cr}
\right. \eqlabel\nbk$$
Further variations on the idea of threshold level are presented in
[\ref{M. Walton, Can. J. Phys.  {\bf 72} (1994) 527.}].

Let us finally note that the fusion coefficients are invariant under the
following action of the outer-automorphism group [\wal]

$$ {\Nc_{A\lah, A'\muh}^{(k)}}^{~AA'\nuh}=
{\Nc_{\lah\muh}^{(k)}}^{~\nuh}\eq$$ 

For example, for $\sp(4)$, the non-trivial outer automorphism $a$ exchanges the
zeroth and second root, or equivalently, it acts on weights as
$a[\la_0, \la_1, \la_2] = [\la_2, \la_1, \la_0]$. 
Acting on the fusion (\ess) as
$$a[0,1,1]\times a[0,1,1] =[1,1,0]\times
[1,1,0]=[2,0,0]\,\oplus\,[1,0,1]\,\oplus\,[0,2,0]
\eq$$
which is easily checked from the tensor product
$$(1,0)\otimes(1,0) = (0,0)\oplus\, (0,1)\oplus \, (2,0)\eq$$
which is non-truncated at level 2.
 Other fusions at level 2 can be obtained from (\ess) by
 acting on the weights as follows
$$\eqalign{ a[0,1,1]\times [0,1,1] =[1,1,0]\times
[0,1,1]&=a[2,0,0]\,\oplus\,a[1,0,1]\,\oplus\,a[0,2,0]\cr &= 
[0,0,2]\,\oplus\,[1,0,1]\,\oplus\,[0,2,0]\cr}\eq$$



The algorithm underlying the Kac-Walton formula suggests a simple road to the
construction of fusion-rule generating functions, that is by starting from the
tensor-product calculation, but keeping track of the level and taking into
account the action of the affine Weyl group. We
illustrate the method for the simple
$\su(2)$ case.
Recall that the $su(2)$ tensor-product 
generating function reads

$$G^{su(2)}(L,M,N)= {1\over 
(1-LM)(1-LN)(1-MN)}\eqlabel\sufun$$
We start with the generating function
$$F(d,L,M,N) = 
 {1\over 
(1-d)(1-LM)(1-LN)(1-MN)}.\eqlabel\fus$$
This is just the generating function for $su(2)$ tensor
products divided by $(1-d)$. The exponent of $d$ will be identified
with the level. We will proceed to the generating function for
$\su(2)$ fusion rules by modifying (\fus). First note that at level $k$
we need only consider the products of $su(2)$ representations $(a)$
with $a\leq k$. The generating function (\fus) includes products of
representations which violate this condition. To keep terms
of the form $d^kL^a$ with $a\leq k$ introduce a dummy variable
$x$ 
(using MacMahon's notation -- cf. [\BCM])
$$
\Oxx \;{1\over{(1-x^{-1})}} F(dx,Lx^{-1},M,N)
\eq$$
This first converts $d^kL^a$ to $x^{-m+k-a}d^kL^a$, with $m\geq 0$ and then keeps
terms of degree zero in $x$ which 
corresponds to keeping the terms of $F(d,L,M,N)$
with $a\leq k$ as required.
This yields:
$$
 {1\over 
(1-d)(1-dLM)(1-dLN)(1-MN)}.\eq$$
Repeating this procedure with $L$ replaced by $M$ yields:
$$G(d,L,M,N) = 
 {1-d^2LMN^2\over 
(1-d)(1-dLM)(1-dLN)(1-dMN)(1-dLMN^2)}.\eq$$
This is still a generating function for tensor products, but with the
size of the representation Dynkin labels restricted to be less than
or equal to the level.

To take into account the affine Weyl group, consider a term in the expansion
of the generating function which contains $d^kN^c$.  If $c\geq k+1$ then this
representation is reflected back into the fundamental region of the 
affine Weyl group: $c\mapsto c - 2(c-k-1) = -c+2k+2$ or 
$d^kN^c\mapsto d^kN^{2k-c+2}$. Since this is a reflection, the corresponding
character must be subtracted. In principle other affine Weyl transformations
might be necessary to obtain a weight in the fundamental domain, but,
as discussed earlier,
for $\su(2)$ one reflection suffices.
At the level of generating functions the effect is to 
replace $$G(d,L,M,N)\quad \mapsto \quad G(d,L,M,N) - N^2G(dN^2,L,M,N^{-1})\eq$$
 Note
that the new generating function contains terms with negative powers
of $N$ and also terms with $c>k$. To obtain the final function we
projected out the required terms as above.
Although this calculation is somewhat long (the verification here was done on a
computer), the final result is very simple:
$$G^{\su(2)}= {1\over
(1-d)(1-dLM)(1-dLN)(1-dMN)}\eqlabel\fgdeux$$
This has first been written down in [\CMW]. There are thus four elementary
couplings:
$$\eqalignQ{ &\E_0 : d~:\quad &(0)\otimes(0)\supset(0)_1\qquad
& \E_2 : dLN: &(1)\otimes(0)\supset(1)_1,\cr
& \E_1 : dLM: &(1)\otimes(1)\supset(0)_1,\qquad  & \E_3 : dMN:
&(0)\otimes(1)\supset(1)_1.\cr}
\eqlabel\sudeuele$$ As
explained above, subscripts indicate the {\it threshold level}.

The notion of a model was discussed in 
[\BCM]. A model for this generating function is  $\Q[\hat E_0,\hat E_1,\hat E_2,\hat
E_3]$ with the gradings of $\hat E_0, \cdots ,
\hat E_3$ respectively given by 
$(1,0,0,0), \,(1,1,1,0), \, (1,1,0,1)$ and $(1,0,1,1)$ for the ordering $X_0=d,\,
X_1=L, X_2=M,\, X_3=N$. As for the finite $su(2)$ case, there are no relations
between the elementary couplings.

The generalisation of the above calculation to other affine Lie algebras is
straightforward.  Starting from the tensor-product generating function augmented by
the factor
$1/(1-d)$, where $d$ keeps track of the level, one  first enforces the integrability
requirement of the first two weights (those that are fused together); one then
implements all the affine reflections of the  set $\W_f$  on the third weight and
projects the alternating sum onto the integrable sector.  However, even though the
strategy is clear, the computations become rapidly very complicated.

To bypass this difficulty, we have
argued (cf. [\BCM] section 3) that the use of
 a direct description of tensor products in terms of a system of inequalities
(e.g., the Littlewood-Richardson (LR) inequalities underlying their combinatorial
description for calculating tensor products -- cf. [\BCM] section 4)  simplifies
the general procedure to a very large extent in addition to allowing us to use
powerful algebraic results.  We now look for a
similar procedure here.  
However, this program faces an immediate difficulty since even for
$su(N)$, a combinatorial description of fusion rules is not known. 
Our method is instead to find an independent route leading
to the elementary couplings. Indeed, the elementary couplings are really what
we need in order to apply our Grobner basis machinery.    Quite remarkably,
it turns out that once elementary couplings  are found, there is a method that
allows us to reconstruct the underlying system of Diophantine inequalities. 

\newsec{Fusion-rule elementary couplings}

The construction of this section depends upon 
the following:

\n {\it Fundamental conjecture}: There exists a fusion basis, that is, a set linear
and homogeneous inequalities involving $k$ and containing as a subset, a
tensor-product basis.

For instance, the LR basis is a set linear and homogeneous inequalities.  Every
solution can be expanded in terms of the elementary solutions of these
inequalities. For $\su(N)$, the conjecture amounts to the existence of a
set of additional inequalities  involving the level $k$ that provide the proper
truncation describing the fusion rules. The relation of this conjecture to the
conjectures presented in [\CMW] is discussed in the Appendix B.

Note that homogeneity  is the key property 
which allows us to reconstruct the fusion
basis from a set of fusion elementary couplings using
Farkas' Lemma.
This condition does not necessarily hold, for
example we have found that the Lie superalgebra $osp(1,2)$  
does not have a homogeneous basis.

Given homogeneity and 
Farkas' lemma, the problem is reduced to finding a set of 
fusion elementary couplings.
The Kac-Walton algorithm is one possible approach, but a
rather difficult one. Instead, we will  introduce a 
simpler approach  based on the outer-automorphism group. Unfortunately, it relies on another conjecture.

Let us start from the set of tensor-product elementary couplings $\{E_i, i\in I\}$
for some set $I$ fixed by the algebra under study. For each $E_i$, we  calculate
the threshold level $k_0(E_i)$.
This information specifies the affine extension of $E_i$. The
affine extension of a tensor-product elementary coupling is necessarily a fusion-rule
elementary coupling
given our hypothesis that the
fusion basis contains, as a subsystem, the set of inequalities that describe tensor
products. Denoting by a hat the affine extension of a tensor-product elementary
coupling
$$\E_i= d^{k_0(E_i)}E_i\eq$$ we have then a partial set of fusion elementary couplings
with the set
$\{\E_i,i\in I\}$. Our conjecture is that the missing fusion elementary couplings
can all be generated by the action of the outer-automorphism group
whenever this group is nontrivial:

\n  {\it The outer-automorphism completeness conjecture}: The complete set of
elementary couplings $\{\E_i,i\in J\}$ for a set $J\supset I$ can be generated by the
action of the outer-automorphism group on the set $\{\E_i,i\in I\}$, i.e., the full
set is contained in
$\{ {\cal A }\E_i\}$: 
$$\{\E_i,i\in J\}\subset \{ {\cal A }\E_i, i\in I\}\eq$$

The 
action of the
outer-automorphism group  on a coupling is defined as follows. Let the three
weights in the coupling be 
$\{\lah,
\muh;
\nuh\}$ where $\nuh\subset \lah\times \muh$, then 
 $${\cal A }\{\lah,
\muh;
\nuh\}= \{A\lah,
A'\muh;
AA'\nuh\}\eq$$ where $A,A'$ are arbitrary elements of the  outer-automorphism
group; the conjectured completeness requires the consideration of all possible
pairs $(A,A')$. 

It should be stressed that we do not suppose that the action of ${\cal A }$ on an
elementary coupling will necessarily produce another elementary coupling.  Indeed,
the resulting coupling could be a product of elementary couplings.  What is
conjectured here is that all fusion elementary couplings can be generated in this
way.

 If the outer-automorphism group is trivial, we
expect that there will a single extra elementary coupling, the one associated to the
scalar coupling: $\E_0$.

As a simple example consider $\su(2)$. Start with the elementary
coupling $E_1:\, (1)\otimes (1)\supset (0)$.
It is easy to show that this coupling arises at level 1. This is thus the
value of its threshold level. The corresponding fusion is
$[0,1]\times [0,1] \supset [1,0]$.
We now consider  all possible actions of the outer-automorphims group on it. 
Since this group is of order 2, there are 4 possible choices for the pair
$$(A,A')\in \{(a,a),(1,1), (1,a), (a,1)\}\eq$$
with 
$a[\la_0, \la_1] = [\la_1, \la_0]$.
This generates the following set of four elementary couplings 
found previously (cf.
eq (\sudeuele)):
$$\eqalignQ{ 
&\E_0 : d~:\quad &[1,0]\times [1,0]\supset[1,0]\qquad
&\E_2 : dLN: &[0,1]\times [1,0] \supset [0,1]\cr
& \E_1 : dLM: &[0,1]\times [0,1] \supset [1,0]\qquad
 & \E_3 : dMN:
&[1,0]\times [0,1] \supset [0,1].\cr}
\eqlabel\sudeuelex$$

Let us then suppose that we have a complete set of fusion elementary
couplings which are the elementary solutions of set of linear and homogeneous
inequalities that we are looking for. 
A standard theorem in the theory of linear Diophantine equations
(cf. [\ref{R.P. Stanley, Duke Math. J. {\bf 40} (1973) 607; 
{\it Combinatorics and Commutative Algebra}, (Boston:
Birkhauser) (1983).}\refname\Stan]) 
states  that every non-negative
integer  solution of a given set of homogeneous  Diophantine inequalities for the
variables
$x_i$ (e.g., for $su(N)$, these are the $\{\la_i, n_{ij}\}$) can be generated from a
non-negative combination of the fundamental solutions.  Hence, given the set of
elementary couplings
$\{\E_i\}$, any coupling can be decomposed (maybe not uniquely) in the form
$\prod_i \E_i^{\,a_i}$.
Let the grading variables representing the $x_i$ be denoted by $X_i$.  To the
expression of $g(\E_i)$ corresponds a vector $\e_i$ of components $\e_{ij}$, which
is the vector form of the elementary solutions of the Diophantine equations. In
other words,
$$\E_i: \, g(\E_i) = \prod_j X_j^{\e_{ij}}\eqlabel\epde$$ Reading off a
particular coupling means that we are interested in a specific set of
non-negative integers $\{x_i\}$ given by
$$\sum_i a_i\e_{ij}  = x_j\eq$$
in terms of non-negative integers $a_i$.  We are thus looking for the existence
conditions for such a coupling. This is related to Farkas' lemma [\ref{A.
Schrijver, {\it Theory of linear and integer programming}, Wiley
1986.}\refname\Schri,\ref{D. Beklemichev, 
{\it Cours de g\'eom\'etrie analytique et
d'alg\`ebre lin\'eaire}, Mir. p. 519.}].
The standard, rational,
form of the lemma is (cf. [\Schri], corollary 7.1d): 

\n {\it Farkas' lemma}: Let  $V$ an $m\times n$ matrix with
rational entries and
let $x\in\Q^m$. Then there  exists
$a\geq 0$, $a\in \Q^n$ such that $Va=x$ if and only if 
for all $u\in \Q^m$, $u^\top V \geq 0$ implies
$u^\top\, x\geq 0$.

We can relate this to our problem in the following way. First note that
the condition that for all $u\in \Q^m$, 
$u^\top V \geq 0$ implies $u^\top\, x\geq 0$
is equivalent to the condition that
for all
$u\in \Z^m$, $u^\top V \geq 0$ implies $u^\top\, x\geq 0$.
Necessity is clear and sufficiency follows since if $u\in \Q^m$ and
 $u^\top V \geq 0$ then $u= cu'$ with $c\in \Q$, $c>0$ and $u'\in\Z^m$.
Then  $u'^\top V \geq 0$, so $u'^\top\, x\geq 0$ and multiplying
by $c$ gives the required inequality.

Now consider the inequalities  
$$u^\top V \geq 0,\quad u\in \Z^m. \eqlabel\condA $$
By writing $u_i=w_i-v_i$,  $w_i,v_i\in\N$, $i=1\dots m$, we obtain a 
new system of linear Diophantine inequalities. It is not difficult
to see that every solution to (\condA) 
can be obtained from a solution to this new system. Moreover,
the new system of linear
Diophantine inequalities has a finite set of fundamental solutions. 
These
give rise to a set of fundamental 
solutions to (\condA) such that every solution
to (\condA) is a linear combination of these fundamental solutions with
non-negative integer coefficients. Call these fundamental solutions
$s_i$, $i=1\dots k$. Thus the condition
that for all $u\in \Z^m$, $u^\top V \geq 0$ implies $u^\top\, x\geq 0$
is equivalent to the condition
$s_i^\top x \geq 0$, $i=1\dots k$.

Putting all this together we obtain the following variation
of Farkas' lemma:

\n {\it Lemma}: Let  $V$ be an $m\times n$ matrix with
rational entries and
let $x\in\Q^m$. Then there  exists
$a\geq 0$, $a\in \Q^n$ such that $Va=x$ if and only if 
$s_i^\top x \geq 0$, $i=1\dots k$ where $s_i$, $i=1\dots k$
are a fundamental set of solutions of the system
$u^\top V \geq 0$, 
$u\in \Z^m$.

%

We can reformulate this Lemma over the integers in a form which
is more convenient for our application:

\n {\it Proposition A}: 
Suppose $V\in M_{m,n}(\N)$ 
and
let $x\in\N^m, a\in \N^n$. Then 
 $Va=x$ 
if and only if 
$u_i^\top x =\alpha_i^\top a $, $i=1\dots k$ where $u_i,\alpha_i$,
$i=1\dots k$
are a fundamental set of solutions of the system
$u^\top{V} = \alpha^\top$, 
$u\in \Z^m$, $\alpha\in\N^n$.

To show this, suppose 
that $Va=x$
and that $u^\top{V} = \alpha^\top$ with
$u\in \Z^m$, $\alpha\in\N^n$. 
Then  $u^\top x = u^\top Va=\alpha^\top a$.
In particular this is true for the fundamental 
solutions.

Conversely, suppose $u_i^\top x =\alpha_i^\top a $ for
every fundamental solution. Then $u^\top x =\alpha^\top a $
for every $u\in \Z^m$, $\alpha\in\N^n$ such that
$u^\top{V} = \alpha^\top$. 
Since $V\in  M_{m,n}(\N)$,  one set of solutions
to $u^\top{V} = \alpha^\top$, $u\in \Z^m$, $\alpha\in\N^n$
is given by taking $u$ to be a suitable unit vector
and $\alpha^\top$ to be a row of $V$ which gives $Va=x$ as required.


To link  the lemma to the situation presented  above, we note that the entries
$V_{ij}$ of the matrix
$V$ are given here by the numbers $\e_{ji}$ appearing in (\epde).
  Our analogue
of the relation
$V\, a=x$ describes a generic coupling and our goal is to find the defining
system of inequalities underlying the existence of this coupling. 
The equalities $u_i^\top x = \alpha_i^\top a$  $i=1\dots k$ imply that $x$
satisfies
$u_i^\top x \geq 0$ $i=1\dots k$
since $\alpha_i$ and $a$ are non-negative.
In general these  inequalities have solutions
which are not solutions of the former equalities for any $a$. 
For example if $V=(2)$, then $Va=x$ is $2a=x$ which is
also the equality obtained from the second part of the
Proposition A. Thus $x$ is a non-negative even integer.
But the corresponding inequality is $x\geq 0$.
However, we have
found that for the particular systems we consider, this does not happen - as can
be easily verified by computing the fundamental set of solutions to the
inequalities
$u_i^\top x \geq 0$ $i=1\dots k$ and verifying that they are the
columns of $V$.




As a simple illustration of this construction, let us work out
 the example of
$\su(2)$.  We use the LR variables
$\{k,
\la_1, n_{11}, n_{12}\}$ and the corresponding grading variables $\{d, L_1,
N_{11}, N_{12}\}$ in terms of which the
elementary couplings and the corresponding vectors are
$$\eqalignD{ &\E_0 : d\qquad & \e_0= (1,0,0,0)\cr
&\E_1: dL_1N_{12}\qquad  & \e_1= (1,1,0,1)\cr
&\E_2: dL_1\qquad  & \e_2= (1,1,0,0)\cr
&\E_3: dN_{11}  & \e_3= (1,0,1,0)\cr}\eqlabel\eleuu$$
For future reference, we display the LR tableaux of the
corresponding tensor-product elementary couplings
$$E_1: \quad \ST{\STrow{\bv}\STrow{\b1}}\, ,  
\qquad E_2: \quad \ST{\STrow{\bv}}\, ,
\qquad E_3: \quad \ST{\STrow{\b1}}\eqlabel\deuxtab$$
To the fusion elementary couplings, we associate the vectors $\e_{j}$ which form
the matrix $V$ with components $V_{ij}=\e_{ji}$:
$$V= \pmatrix {1&1&1&1\cr 0&1&1&0\cr 0&0&0&1\cr 0&1&0&0\cr}\eq$$
and so we have the matrix equation
$$V\, a=x\eq$$
This equation describes a general fusion coupling.  We now want to unravel the
underlying system of inequalities.  For this, we 
use Proposition A, i.e., we find the fundamental
solutions of 
$u^\top\, V\geq 0$.
This is first transformed into a set of equalities $u^\top\, V= \a^\top $ by
introducing new non-negative parameters $\a_i$:
$$\eqalignD{& u_0= \a_0\qquad\qquad &u_0+u_1= \a_2\cr
& u_0+u_1+u_3= \a_1\qquad\qquad
&u_0+u_2= \a_3\cr}\eq$$
We next apply the vector-basis arguments (see Section 7 of [\BCM]).
Let us choose the $\a_i$ as our independent variables.  (This example is somewhat
misleading due to its simplicity: in general not all the $\a_i$ can be taken as
the independent variables.) The dependent variables read then
$$\eqalignD{& u_0= \a_0\qquad\qquad &u_2= \a_3-\a_0\cr
& u_1= \a_2-\a_0\qquad\qquad
&u_3= \a_1-\a_2\cr}\eq$$
The 4 basis  vectors are obtained by setting successively one $\a_i$ equal to 1 and
all the others equal to 0.  These vectors are written as $e_i$ and their entries are 
$$e_i= \left(u_0(\a_i=1),\,u_1(\a_i=1),\,u_2(\a_i=1),\,u_3(\a_i=1); \a_0,
\a_1,\a_2,\a_3\right)
\eq$$ With $i=0,1,2,3$, we find
$$\eqalignD{& e_0=(1,-1,-1,0;1,0,0,0)\qquad\qquad &e_3=(0,0,1,0;0,0,0,1)\cr 
& e_1=(0,0,0,1;0,1,0,0)\qquad\qquad & e_2=(0,1,0,-1;0,0,1,0)
\cr}\eq$$ These $e_i$ are manifestly linearly independent
and they are non-negative expressions in the $\a_i$. In other words, their
grading re-transcription of the above vectors (with $U_i$ and $\Ac_i$ denoting
the grading variables of $u_i$ and $\a_i$ respectively) reads
$$\eqalignD{& \Ec_0=U_0U_1^{-1}U_2^{-1}\Ac_0\qquad\qquad & 
\Ec_2=U_1U_3^{-1}\Ac_2\cr
&\Ec_1=U_3\Ac_1\qquad\qquad 
&  \Ec_3=U_2\Ac_3\cr}\eqlabel\drdr$$ 
Here we see that all $\Ec_i$ contain positive powers of the $\Ac_i$ (this is not
generic and it reflects the simplicity of the $su(2)$ case).  Hence, all solutions
are generated freely from the non-negative powers of the
$\Ec_i$.

The corresponding linear system of Proposition A 
is $e_i(x,-a)^\top = 0$ with
$x=(k,\la_1,n_{11},n_{12})$ and $a=(a_1,a_2,a_3,a_4)$ non-negative integers:
$$\eqalignD{ & k- \la_1-n_{11}=a_1\qquad\qquad &\la_1- n_{12}=a_3\cr
&n_{12}= a_2\qquad\qquad
&n_{11}=a_4\cr}\eqlabel\inedeux$$
which are equivalent to the inequalities:
$$\eqalignD{ & k\geq \la_1+n_{11}\qquad\qquad &\la_1\geq n_{12}\cr
&n_{12}\geq 0\qquad\qquad
&n_{11}\geq 0\cr}\eqlabel\inedeux$$
The last three conditions define the LR basis.  The first one is the additional
fusion constraint.

In general, we will work the elementary solutions $e_i$ in their exponential
version $\Ec_i$ to keep the notation more compact and it should be clear that the
(in)equalities can be read off as easily at this level.

 The construction of the $\su(2)$ generating function  is now
straightforward: since
there are no  relations between the elementary couplings, the generating function is
simply (\fgdeux), that is
$$G^{\su(2)} = \prod_{i=0}^3{1\over (1-\E_i)}\eq$$

 From the $k$--inequality of the $\su(2)$ fusion basis, 
we read off the threshold level of a coupling as
$k_0= \la_1+n_{11}$, 
that is
$$k_0=  (\la_1+\mu_1+\nu_1)/2\eqlabel\deuxseuil$$

The threshold level is also nicely coded in the LR tableaux: all elementary couplings
have threshold level 1 and they all have a single column. We can then write
directly that
$$ k_0 = \# {\rm columns} = \la_1+n_{11}\eq$$
and we recover the previous result. For an $su(2)$ LR tableau, it is clear that
the number of columns is given by this expression.  More generally, for $su(N)$, it
is simple to check that the number of columns is simply $$ \# {\rm columns} =
(\la+\mu+\nu, \omega_{N-1}) = \sum_{i=1}^{N-1} \la_i +n_{11}\eq$$
where $\omega_{N-1}$ is the $N-1$-th fundamental weight.




\newsec{The generating function for $\su(3)$ fusion rules}

The $su(3)$ tensor-product elementary
couplings are:
$$\eqalign{ 
  E_1 &=\ST{\STrow{\bv}\STrow{\b1}\STrow{\b2} }, \quad  
 E_2=\ST{\STrow{\bv}}, \quad
 E_3=\ST{\STrow{\b1}}, \quad
 E_4=\ST{\STrow{\bv}\STrow{\bv}\STrow{\b1}} \cr
   E_5 & =\ST{\STrow{\bv}\STrow{\bv}}, \quad
  E_6 =\ST{\STrow{\b1}\STrow{\b2}}, \quad
  E_7 =\ST{\STrow{\bv}\STrow{\b1}}, \quad
  E_8 =\ST{\STrow{\bv\b1}\STrow{\bv}\STrow{\b2}}.
 \cr} \eq$$

Using the Kac-Walton formula, the threshold
level of $E_1$
is 1 and the corresponding fusion reads
$$\E_1: \quad [0,1,0]\times [0,0,1]\supset
[1,0,0]\eq$$
Acting on $\E_1$ with $(a^n,a^m;a^{n+m})$ 
$n,m=0,1,2$ yields the  elementary couplings:
$$\eqalignD{
\E_0: \quad[1,0,0]\times [1,0,0]\supset [1,0,0]:& \quad d \quad &(1,0,0,0,0,0,0,0)
\cr
\E_1:\quad[0,1,0]\times [0,0,1]\supset [1,0,0]: & \quad
dL_1N_{12}N_{23}\quad &(1,1,0,0,1,0,0,1) \cr 
\E_2:\quad[0,1,0]\times [1,0,0]\supset [0,1,0]: & \quad 
dL_1\quad &(1,1,0,0,0,0,0,0) \cr 
\E_3:\quad[1,0,0]\times [0,1,0]\supset [0,1,0]: & \quad 
dN_{11}\quad &(1,0,0,1,0,0,0,0) \cr
\E_4:\quad[0,0,1]\times [0,1,0]\supset [1,0,0]: & \quad
 dL_2N_{13}\quad  &(1,0,1,0,0,1,0,0) \cr 
\E_5:\quad[0,0,1]\times [1,0,0]\supset [0,0,1]: & \quad 
d L_2\quad &(1,0,1,0,0,0,0,0) \cr
\E_6:\quad[1,0,0]\times [0,0,1]\supset [0,0,1]: & \quad 
dN_{11}N_{22}\quad &(1,0,0,1,0,0,1,0) \cr 
\E_7:\quad[0,1,0]\times [0,1,0]\supset [0,0,1]: & \quad 
d L_1N_{12}\quad &(1,1,0,0,1,0,0,0) \cr 
\E_8:\quad[0,0,1]\times [0,0,1]\supset [0,1,0]: & \quad 
d L_2N_{11}N_{23}\quad &(1,0,1,1,0,0,0,1) \cr }\eq$$
The last column is the vector $\e_i$ with entries 
$ (k,\la_1,
\la_2, n_{11}, n_{12}, n_{13}, n_{22}, n_{23})$.  
By this procedure, we have thus recovered the affine extension of the 8
tensor-product elementary couplings and found an extra elementary coupling: $\E_0$.

To derive the fusion basis, we proceed as in the $su(2)$ case.
The set of variables here is
$$(x_0, x_1,\cdots, x_7) = (k,\la_1,
\la_2, n_{11}, n_{12}, n_{13}, n_{22}, n_{23})\eq$$
and the matrix $V$ (with columns written in the order $\E_0, \cdots,\E_8$) reads
$$V= \pmatrix {1&1&1&1&1&1&1&1&1\cr
0&1&1&0&0&0&0&1&0\cr
0&0&0&0&1&1&0&0&1\cr
0&0&0&1&0&0&1&0&1\cr
0&1&0&0&0&0&0&1&0\cr
0&0&0&0&1&0&0&0&0\cr
0&0&0&0&0&0&1&0&0\cr
0&1&0&0&0&0&0&0&1\cr}\eq$$
The reformulation of $u^\top\, V\geq 0$ in terms of
equalities by the introduction of appropriate nonnegative parameters
reads:
$$\eqalignD {
&u_0= \a_0\qquad &u_0+u_2= \a_5\cr
&u_0+u_1+u_4+u_7= \a_1\qquad &u_0+u_3+u_6= \a_6\cr
&u_0+u_1= \a_2\qquad &u_0+u_1+u_4= \a_7\cr
&u_0+u_3= \a_3\qquad &u_0+u_2+u_3+u_7= \a_8\cr
&u_0+u_2+u_5= \a_4\qquad ~ \cr
}\eq$$
We have 17 variables and 9 equations, hence 8 free variables.  Let us choose them
to be the $\a_i$ except for $\a_5$.  Solving for the dependent variables leads to
$$\eqalignD{
&u_0= \a_0\qquad &u_5= -\a_0+\a_1+\a_3+\a_4-\a_7-\a_8\cr
&u_1= -\a_0+\a_2\qquad &u_6= -\a_3+\a_6\cr
&u_2= -\a_1-\a_3+\a_7+\a_8\qquad &u_7= \a_1-\a_7\cr
&u_3= -\a_0+\a_3\qquad &\a_5= \a_0-\a_1-\a_3+\a_7+\a_8\cr
&u_4= -\a_2+\a_7\qquad ~\cr}\eqlabel\sstt$$
The basis vectors $e_i$ of this system are obtained by setting one of the $\a_j=1$
and all the others equal to 0 (with the understanding the $\a_5$ is excluded from
this list of free variables).  It appears more natural here to express them in their
exponentiated version since a projection will be needed to extract the
non-negative 
fundamental solutions.  Denote by 
$U_i$ the grading variable associated to
$u_i$ and by $\Ac_i$ those associated to
$\a_i$, the exponential form of the basis vectors reads 
$$\eqalignD{
&\Ec_0: U_0U_1^{-1}U_3^{-1}U_5^{-1}\Ac_0 \Ac_5 \qquad  &\Ec_4: U_5 \Ac_4\cr
&\Ec_1: U_2^{-1}U_5U_7 \Ac_1\Ac_5^{-1}\qquad  &\Ec_5: U_6 \Ac_6\cr
&\Ec_2: U_1U_4^{-1} \Ac_2\qquad &\Ec_6: U_2U_4U_5^{-1}U_7^{-1}\Ac_5 \Ac_7\cr
&\Ec_3: U_2^{-1}U_3U_5U_6^{-1}\Ac_3\Ac_5^{-1}\qquad &\Ec_7: U_2U_5^{-1}\Ac_5
\Ac_8\cr}\eq$$
To get the corresponding non-negative couplings, i.e., terms containing only non-negative
powers of the $\Ac_i$, we must keep only the non-negative powers of the $\Ec_i$.
But this is not sufficient since negative powers of $\Ac_5$ can appear:  we need to
project the free generators of the non-negative $\Ec_i$ powers 
$$\prod_{i=0}^7 {1\over 1-\Ec_i}\eq$$
to non-negative $\Ac_5$ powers, using, say the MacMahon algorithm (cf. the
$\Omega$ projection described in section 3 of [\BCM]). After the projection,
all the variables $\Ac_i$ are set equal to 1.  Here however, it is fairly easy to
find out by inspection those non-negative combinations of the $\Ec_i$ that have
non-negative
$\Ac_5$ terms. These are
$$\Ec_0, \Ec_2, \Ec_4, \Ec_5, \Ec_6, \Ec_7\eq$$ together with
$$\eqalignD{
& \Ec_0\Ec_1: U_0U_1^{-1}U_2^{-1}U_3^{-1}U_7 \Ac_0 \Ac_1
\qquad & \Ec_1\Ec_7: U_7\Ac_1 \Ac_8\cr 
& \Ec_0\Ec_3: U_0U_1^{-1}U_2^{-1}U_6^{-1}\Ac_0 \Ac_3\qquad  
& \Ec_3\Ec_6: U_3U_4U_6^{-1}U_7^{-1} \Ac_3 \Ac_7\cr
& \Ec_1\Ec_6: U_4 \Ac_1
\Ac_7\qquad & \Ec_3\Ec_7: U_3U_6^{-1} \Ac_3 \Ac_8\cr}\eq$$

At this point, we set all $\Ac_i=1$. We have thus 12 elementary 
non-negative solutions and the corresponding inequalities are:
$$\eqalignT{ &\la_1 \geq n_{12}\qquad  &\la_2 \geq n_{13}\qquad 
&\la_2+n_{12} \geq n_{13}+n_{23}\cr &n_{11}\geq
n_{22}\qquad
  &n_{11}+n_{12}\geq n_{22}+n_{23}&~\cr}
\eqlabel\inee$$
and $n_{ij}\geq 0$ (except for $n_{11}\geq0$ which is implied by the others),
which are the LR conditions for $su(3)$.  There are also three
inequalities involving
$k$:
$$\eqalign{ k-\la_1-\la_2 &\geq n_{22}\cr
  k-\la_1-\la_2 &\geq n_{11}-n_{23}\cr  k-\la_1 &\geq n_{13}+n_{11}\cr}
\eqlabel\ineee$$
The set of inequalities (\inee) and (\ineee) represents the $\su(3)$ fusion basis.


Before we leave the analysis of the $\su(3)$ case, let us return to the set of
equations (\sstt).  The last equality gives a relation between different $\a_i$.
Actually this relation signals a relation between different sums of columns of $V$.
In other words, this signals a relation between products of elementary couplings. 
Indeed, to link the last equality of (\sstt) with such a relation, we recall that
the labelling of the $\a_i$ is that of the elementary couplings, which are the
columns of $V$.  Hence, the sought for relation is simply the product form of the
equality with
$\a_i\mapsto\E_i$:
$$\a_1+\a_3+\a_5= \a_0+\a_7+\a_8 \quad \mapsto \quad \E_1\E_3\E_5=
\E_0\E_7\E_8\eq$$

As there  is only one relation, it is easy to find the generating function.  Forbidding
$\E_1\E_3\E_5$, we get [\CMW]
$$\eqalign{ {G_1}&=\left(\prod_{i=0\atop i\not=1,3,5}^8~(1-\E_i)^{-1}\right)
\left({1
\over (1-\E_1) (1-\E_5)} \right.\cr & \left.\qquad\qquad +{\E_3  \over (1-\E_3)
(1-\E_1)}+{\E_3\E_5
\over (1-\E_5) (1-\E_3)}\right)\cr}\eqlabel\fussa$$
If instead, we
decide to forbid $\E_0\E_7\E_8$, we would have
$$\eqalign{{G'} &=\left(\prod_{i=0\atop i\not=0,7,8}^8~(1-\E_i)^{-1}\right)
\left({1 \over (1-\E_0) (1-\E_7)} \right.\cr & \left.\qquad\qquad +
{\E_8  \over (1-\E_7) (1-\E_8)}+
{\E_0\E_8\over (1-\E_8) (1-\E_0)}\right)\cr}\eqlabel\fussb$$
and simple manipulations show that ${G_1}= {G'}$.
An
independent proof of this generating function is presented in Appendix A.

Given the fusion basis, we can write down directly the threshold level to be
$$k_0= {\rm max} (\la_1+\la_2+n_{11}-n_{23},\,  \la_1+\la_2+n_{22},
\,\la_1+n_{11}+n_{13})\eqlabel\troisseuil$$
This can also be extracted from the generating function as follows.
A generic term of
the $\su(3)$ generating function is ($\E_0=d$)
$$d^\alpha\E_1^a\E_2^b\E_3^c\E_4^d\E_5^e\E_6^f\E_7^g\E_8^h\eq$$
with either $a=0,\, c=0$ or $e=0$. In all cases the threshold level is simply
$$k_0 = a+b+c+d+e+f+g+h\eq$$
In terms of the grading variables
$L_i$ and $N_{ij}$, the above generic term becomes
$$d^{\alpha+k_0}L_1^{a+b+g}L_2^{d+e+h}N_{11}^{c+f+h}
N_{12}^{a+g}N_{13}^{d}N_{22}^{f}N_{23}^{a+h}\eq$$
{}From this expression we read off the relation between the $n_{ij}$ and the
variables $a,\cdots, h$. In each three cases (where one of $a,c,e$ is zero), we
can then solve for the sum $k_0 = a+b+c+d+e+f+g+h$. We find
$$\eqalignD{ & a= 0:  \quad k_0 &=~\la_1+\la_2+n_{11}-n_{23}\cr
& c= 0:  \quad k_0&=~\la_1+\la_2+n_{22}\cr
& e= 0:  \quad k_0&=~ \la_1+n_{11}+n_{13}\cr}\eq$$
This leads to the compact expression (\troisseuil) for the $\su(3)$ threshold level.
This is easily checked to be equivalent to the formula given in [\KMSW,\BKMW] in
terms of BZ triangle data (cf. section 7.1 of [\BCM]):
$$k_0=\max\{m_{13}+\mu_1+\mu_2,n_{13}+\nu_1+\nu_2,l_{13}+\lambda_1+\lambda_2\}
\eqlabel\mme$$
An explicit formula for the $\su(3)$ fusion coefficients is written
down in [\ref{L. B\'egin, P. Mathieu
and M.A. Walton, Mod. Phys. Lett. A, Vol. 7
 (1992) 3255.}\refname\LB].

Notice  that  the threshold level is also simply encoded in the LR tableaux.
Indeed, every elementary couplings has threshold level 1 and it corresponds to the
number of columns except for
$E_8$.  This leads  directly to the following formula for the threshold level of a
general LR tableau
$$k\geq k_0\equiv \# \,{\rm columns} - \# E_8 =  \#\,{\rm columns} -
\# \ST{\STrow{\bv\b1}\STrow{\bv}\STrow{\b2}}\eqlabel\tabko$$ that is, $k_0$ is the
number of columns minus the total number of
$E_8$ that we can take out of the tableau while preserving its LR character.
Consider for instance:
$$\ST{\STrow{\bv\bv\bv\b1}\STrow{\bv\b1\b2}\STrow{\b2}}: \qquad
\ST{\STrow{\bv\bv\bv\b1}\STrow{\bv\b1\b2}\STrow{\b2}} - 
 \ST{\STrow{\bv\b1}\STrow{\bv}\STrow{\b2}}= 
\ST{\STrow{\bv\bv}\STrow{\b1\b2}}\eq$$
After the subtraction of one $E_8$, the resulting tableau is not a LR tableau:
counting from right to left, we find that a $\ST{\STrow{\b2}}$ precedes the first 
$\ST{\STrow{\b1}}$. Therefore, no $E_8$ can be removed and $k_0$ is given by the
number of columns which is 4.



\newsec{The $\sp(4)$ generating function}

We first recall some results obtained in [\BCM]. The  appropriate basis for the
description of
$sp(4)$ tensor products reads [\ref{ A.D. Berenstein and
A.V.  Zelevinsky, J. Geom. Phys. {\bf 5} (1989) 453.}\refname\BZin]:
$$
\eqalignD{&\lambda_1  \geq p \qquad &\mu_1  \geq q \cr
&\lambda_2  \geq r_{1}/2   \qquad &\mu_1  \geq q+r_1-r_{2} \cr &\lambda_2  \geq
r_{1}/2+q-p
 \qquad & \mu_1  \geq p+r_1-r_{2} \cr &\lambda_2  \geq r_{2}/2+q-p
\qquad & 
\mu_2  \geq r_{2}/2 \cr
&\nu_1  = r_2-r_1-2p+\la_1+\mu_1\; \qquad &
\nu_2  = p-q-r_2+\la_2+\mu_2\cr}
 \eqlabel\bzspq $$
together with $p, q\in \NN$ and $r_i\in 2\NN$ for $i=1,2$.
A proper set of variables for a complete description of a particular
tensor-product coupling is thus
$$\{\la_1,\la_2,\mu_1,\mu_2, r_1,r_2,p,q\}\eq$$ (notice the absence of the
$\nu_i$ Dynkin labels).  Let the corresponding grading  variables be
$$\{L_1,L_2,M_1,M_2,R_1,R_2,P,Q\}\eq$$ The list of elementary coupling with their
grading description is:
$$\eqalignT{
&A_1:  &(0,0)\otimes(1,0)\supset(1,0)\quad M_1\cr
&A_2:  &(1,0)\otimes(0,0)\supset(1,0)\quad L_1\cr
&A_3:  &(1,0)\otimes(1,0)\supset(0,0)\quad L_1M_1PQ\cr
&B_1:  &(0,0)\otimes(0,1)\supset(0,1)\quad M_2\cr
&B_2:  &(0,1)\otimes(0,0)\supset(0,1)\quad L_2\cr
&B_3:  &(0,1)\otimes(0,1)\supset(0,0)\quad L_2M_2R_1^2R_2^2\cr
&C_1:  &(0,1)\otimes(1,0)\supset(1,0)\quad L_2M_1Q\cr
&C_2:  &(1,0)\otimes(0,1)\supset(1,0)\quad L_1M_2R_2^2P\cr
&C_3:  &(1,0)\otimes(1,0)\supset(0,1)\quad L_1M_1P\cr
&D_1:  &(2,0)\otimes(0,1)\supset(0,1)\quad L_1^2M_2R_2^2P^2\cr
&D_2:  &(0,1)\otimes(2,0)\supset(0,1)\quad L_2M_1^2R_1^2\cr
&D_3:  &(0,1)\otimes(0,1)\supset(2,0)\quad L_2M_2R_2^2\cr}
\eqlabel\spele$$ The
relation between elementary couplings are  generated by
$$\eqalignT{ &C_1 C_2 = A_3 D_3,\quad &C_2C_3= A_1D_1\quad & C_3C_1= A_1A_3B_2
\cr &D_1 D_2=B_3C_3^2\quad & D_2D_3= A_1^2B_2B_3\quad & D_1D_3= B_2C_2^2 \cr
&C_1D_1= A_3B_2C_2\quad &C_2D_2= A_1B_3C_3\quad & C_3D_3= A_1B_2C_2\cr}
\eqlabel\zysp$$

To find the fusion elementary couplings, we start by computing the threshold level
of $A_1$ by the Kac-Walton formula.  It is found to be 1.  The corresponding level-1
fusion, denoted $\A_1$, is thus
$$[1,0,0]\times [0,1,0]\supset [0,1,0]\eq$$
We can act on it with the four pairs $$(A,A')= \{ (1,1), (a,a), (a,1),
(1,a)\}\eq$$ We obtain in this way two copies of $\A_1$ and two copies of $\C_1$,
the level-1 extension of $C_1$.  Similarly, $A_2,$ and $A_3$ are found to have level
1 and this implies the same result for $C_2,C_3$. $B_1$ is also found to have
threshold level 1.  Acting on it with the above sequence of outer automorphisms
leads successively to $B_1, B_2, B_3$ and a new coupling, $\E_0$:
$$\E_0: \quad [1,0,0]\times [1,0,0]\supset [1,0,0]\eq$$
Finally, $D_1,D_2$ and $D_3$ have threshold level 2 and they are all fixed with
respect to the action of the outer-automorphism group.  The set $\{\A_i, \B_i, \C_i,
\D_i, \E_0\}$ is thus our candidate complete set of fusion elementary couplings,
whose explicit expression in terms of grading variables is read from their
tensor-product relative with the addition of an appropriate factors of $d$.  


Having obtained the fusion elementary couplings, we now work out the
corresponding fusion basis.
Introduce the set of variables
$$(x_0, x_1,\cdots, x_8)= 
(k,\la_1,\la_2,\mu_1,\mu_2, r_1,r_2,p,q)\eq$$ This
fixes the ordering of the rows of $V$. The matrix $V$ is built from the columns which
form the different elementary couplings in the order $\E_0, \A_1, \cdots,
\D_3$:
$$V= \pmatrix {
1&1&1&1&1&1&1&1&1&1&2&2&2\cr
0&0&1&1&0&0&0&0&1&1&2&0&0\cr
0&0&0&0&0&1&1&1&0&0&0&1&1\cr
0&1&0&1&0&0&0&1&0&1&0&2&0\cr
0&0&0&0&1&0&1&0&1&0&1&0&1\cr
0&0&0&0&0&0&2&0&0&0&0&2&0\cr
0&0&0&0&0&0&2&0&2&0&2&0&2\cr
0&0&0&1&0&0&0&0&1&1&2&0&0\cr
0&0&0&1&0&0&0&1&0&0&0&0&0\cr
}\eq$$  The transcription of the inequalities $u^\top\, V\geq 0 $ into the
equalities 
$u^\top\, V= \a^\top$ takes the following form
$$\eqalignD{
&u_0=\a_0\qquad &u_0+u_2+u_3+u_8=\a_7\cr
&u_0+u_3=\a_1\qquad &u_0+u_1+u_4+2u_6+u_7=\a_8\cr
&u_0+u_1=\a_2\qquad &u_0+u_1+u_3+u_7=\a_9\cr
&u_0+u_1+u_3+u_7+u_8=\a_3\qquad &2u_0+2u_1+u_4+2u_6+2u_7=\a_{10}\cr
&u_0+u_4=\a_4\qquad &2u_0+u_2+2u_3+2u_5=\a_{11}\cr
&u_0+u_2=\a_5\qquad &2u_0+u_2+u_4+2u_6=\a_{12}\cr
&u_0+u_2+u_4+2u_5+2u_6=\a_6\qquad ~\cr
}\eq$$
Solving for the dependent variables $u_i, \a_j$, $i=0,\cdots, 8$ and $j=6,7,8,9$
gives
$$\eqalignD{
&u_0= \a_0\qquad &u_7= \frac12(\a_0-2\a_2+\a_5+\a_{10}-\a_{12})\cr
&u_1= -\a_0+\a_2\qquad &u_8= \frac12(\a_0-2\a_1+2\a_3-\a_5-\a_{10}+\a_{12})\cr
&u_2= -\a_0+\a_5\qquad &\a_6= -2\a_1-\a_5+\a_{11}+\a_{12}\cr
&u_3= -\a_0+\a_1\qquad &\a_7= \frac12(-\a_0+2\a_3+\a_5-\a_{10}+\a_{12})\cr
&u_4= -\a_0+\a_4\qquad &\a_8= \frac12(-\a_0-\a_5+\a_{10}+\a_{12})\cr
&u_5= \frac12(\a_0-2\a_1-\a_5+\a_{11})\qquad
&\a_9=\frac12(-\a_0+2\a_1+\a_5+\a_{10}-\a_{12})\cr
 &u_6= \frac12(-\a_4-\a_5+\a_{12})\qquad ~\cr
}\eqlabel\ssrr$$
As usual, the basis vectors $e_i$ of this system are obtained by setting one of the
$\a_i=1$ and all the others equal to 0, excluding $\a_6, \cdots ,\a_9$.  We will give
their exponentiated version, where as before, we denote by 
$U_i$ the grading variable associated to
$u_i$ and by $\Ac_i$ those associated to
$\a_i$:
$$\eqalign{
&\Ec_0: U_0U_1^{-1}U_2^{-1}U_3^{-1}U_4^{-1}U_5^{1/2} U_7^{1/2}U_8^{1/2}\Ac_0
\Ac_7^{-1/2}\Ac_8^{-1/2}\Ac_9^{-1/2}\cr 
&\Ec_1: U_3U_5^{-1}U_8^{-1} \Ac_1\Ac_6^{-2}\Ac_9\cr
&\Ec_2: U_1U_7^{-1} \Ac_2\cr
&\Ec_3: U_8\Ac_3\Ac_7\cr
&\Ec_4: U_4 U_6^{-1/2}\Ac_4\cr
&\Ec_5: U_2U_5^{-1/2}U_6^{-1/2} U_7^{1/2}U_8^{-1/2} \Ac_5
\Ac_6^{-1}\Ac_7^{1/2}\Ac_8^{-1/2}\Ac_9^{1/2}\cr 
&\Ec_6:
U_7^{1/2}U_8^{-1/2} \Ac_7^{-1/2}\Ac_8^{1/2}\Ac_9^{1/2}\Ac_{10}\cr
 &\Ec_7:
U_5^{1/2}\Ac_6 \Ac_{11}\cr
&\Ec_8: U_6^{1/2} U_7^{-1/2}U_8^{1/2}
\Ac_6\Ac_7^{1/2}\Ac_8^{1/2}\Ac_9^{-1/2}\Ac_{12}\cr }\eqlabel\sss$$
Next we keep only those combinations of the $\Ec_i$ that contain only 
non-negative integer powers of the
$\Ac_i$.   This projection is not so simple to work out by inspection. We thus need
to use a more systematic procedure:

Consider a general expansion of the form $\prod_i(1-\Ec_i)^{-1}$ and in a generic
term of the form  $\prod_i\Ec_i^{\e'_i}$, let us collect 
the number of ${\cal A}_i$ factors (denote by $a_i$ their exponents).  Of course we
are only interested in those $\Ac_i$ that appear with negative powers, namely
$i=6,7,8,9$.  Their powers can be read off from the  
${\cal A}_i$ in (\sss) and this yields
 the following expressions:
$$\eqalign{
& a_{6}= -2\e'_1-\e'_5+\e'_7+\e'_8\geq 0\cr
& 2a_{7}= -\e'_0+2\e'_3+\e'_5-\e'_6+\e'_{8} \geq 0\cr
& 2a_{8}= -\e'_0-\e'_5+\e'_6+\e'_{8}  \geq 0\cr
& 2a_{9}= -\e'_0+2\e'_1+\e'_5+\e'_{6}-\e'_{8}  \geq 0\cr}\eq$$
(These equations should be compared with the last four of (\ssrr), with $\a_i\rw
\e_i', \, i\leq 5$ and $\a_i\rw e_{i-4}'$ for $i \geq 10$).
 We then look for the elementary solutions of this system of
inequalities.  There are 4 elementary solutions with
$\e'_0\not=0$.  Their grading reformulation reads
$$\Ec_0\Ec_1\Ec_6\Ec_8^2 \qquad\Ec_0\Ec_1\Ec_7\Ec_8 \qquad\Ec_0\Ec_5\Ec_6\Ec_8
\qquad\Ec_0\Ec_3\Ec_6 \eq$$ 
Denote their vector-reformulation respectively as $e_i$ with $i=0,1,2,3$, then the
conditions $e_i \, x\geq 0$ yield, in the above order
$$\eqalign{
&k\geq \la_1+\la_2+\mu_2+r_1/2-r_2\cr
&k\geq\la_1+\la_2+\mu_2-r_2/2\cr
&k\geq \la_1+\mu_1+\mu_2-p\cr
&k\geq \la_1+\la_2+\mu_1+\mu_2-p-q-r_1/2\cr}\eqlabel\trespbas$$
The other elementary solutions are
$$\eqalign{& \Ec_2,\quad\Ec_3,\quad\Ec_4,\quad\Ec_7,\quad\Ec_5\Ec_8,\quad
\Ec_6\Ec_8,\quad \Ec_1\Ec_8^2,\quad 
\Ec_1\Ec_7^2, \cr& \Ec_3\Ec_6^2,\quad
\Ec_5\Ec_6\Ec_7,\quad \Ec_5\Ec_6^2\Ec_8,\quad \Ec_1\Ec_6^2\Ec_8,\quad
\Ec_1\Ec_6\Ec_7\Ec_8\cr}\eq$$
and the resulting inequalities reproduce the whole set of BZ inequalities (\bzspq)
with the positivity requirement on $r_i, p$ and $q$ (together with
$\mu_1\geq q+\frac12(r_1-r_2)$ which is implied by the other ones). 

Let us return to the last four equations in (\ssrr).  As mentioned in connection to
the $\su(3)$ case, they indicate the `basic relations': the correspondence between
the $\a_i$ and the elementary couplings being fixed by the ordering of the
columns of $V$ (e.g., $\a_3\mapsto \A_3$ and $\a_7\mapsto \C_1$).  The relations
correspond then respectively to
$$\eqalignD{ & \A_1^2\B_2\B_3= \D_2\D_3\qquad &\E_0\C_1^2\D_1= \A_3^2\B_2\D_3\cr
 & \E_0\B_2\C_2^2= \D_1\D_3\qquad &\E_0\C_3^2\D_3= \A_1^2\B_2\D_1\cr}\eq$$
The first and  third  relations appear in the list (\zysp). All other
linear relations in the set (\zysp) can be obtained from products of the above four,
allowing for the cancellations of common factors.  For instance, consider the
product of the left factors of the second and third relations; equating this with
the product of the right factors yields
$$\E_0^2\B_2\C_1^2\C_2^2\D_1= \A_3^2\B_2\D_1\D_3^2\eq$$
Cancelling the $\B_2\D_1$ terms and taking the square root gives $$\E_0\C_1\C_2=
\A_3\D_3\eq$$ which is the affine extension of the relation $C_1C_2= A_3D_3$.  All
other linear relations can be obtained in a similar way. 
 


We can write the $\sp(4)$ generating function in the compact form
$$\eqalign{
G &= \Eb_0\,
\Bb_1\Bb_2\Bb_3[\Ab_1\Ab_2\Ab_3\Cb_1\Cb_2\Cb_3(1-\A_1\A_3\B_2)+\D_1\Db_1\Ab_2\Ab_3\Cb_2\Cb_3\cr
&+\D_3\Db_3\Ab_1\Ab_2\Cb_1\Cb_2+ \D_2\Db_2 \Ab_1\Ab_2\Ab_3\Cb_1\Cb_3
(1-\A_1\A_3\B_2)]\cr}\eq$$
where ${\overline Q}$ is defined as
 $${\overline Q} = {1\over 1-{\widehat Q}}\eq$$
This can be re-expressed under a manifestly positive form as follows
$$\eqalign{ G &=
\Eb_0\Ab_1\Ab_2\Bb_1\Bb_2\Bb_3\Cb_1\Cb_2\Cb_3+ 
\Eb_0\A_3\Ab_2\Ab_3\Bb_1\Bb_2\Bb_3\Cb_1\Cb_2\Cb_3\cr
&+ \Eb_0\A_1\A_3\Ab_1\Ab_2\Ab_3\Bb_1\Bb_3\Cb_1\Cb_2\Cb_3+
\Eb_0\D_1\Ab_2\Ab_3\Bb_1\Bb_2\Bb_3\Cb_2\Cb_3\Db_1\cr &+
\Eb_0\D_3\Ab_1\Ab_2\Bb_1\Bb_2\Bb_3\Cb_1\Cb_2\Db_3+
\Eb_0\D_2\Ab_1\Ab_2\Bb_1\Bb_2\Bb_3\Cb_1\Cb_3\Db_2\cr &+
\Eb_0\A_3\D_2\Ab_2\Ab_3\Bb_1\Bb_2\Bb_3\Cb_1\Cb_3\Db_2+
\Eb_0\A_1\A_3\D_2\Ab_1\Ab_2\Ab_3\Bb_1\Bb_3\Cb_1\Cb_3\Db_2\cr}\eqlabel\genfusp$$
We should stress that this is essentially a new result.
A generating function for $\sp(4)
$ fusion rules was given in [\ref{L. B\'egin, P. Mathieu and M.A. Walton, J. Phys. A: Math. Gen. {\bf 25} (1992)
135.}\refname\BMW]; the approach, however, was ad hoc
and the result was not related to any known basis.

As before the information concerning the threshold level  that can be deduced
from   the fusion
basis inequalities (\trespbas) can also be obtained directly from the generating
function.
A generic term of the
$\sp(4)$ generating function (\genfusp) reads
$$d^\alpha \A_1^a\A_2^b\A_3^c \B_1^d\B_2^e\B_3^f \C_1^g\C_2^h\C_3^i
\D_1^j\D_2^k\D_3^l\eq$$
Its threshold level is (since all these factors have a single power of $d$
except for the three $\D_i= d^2D_i$):
$$k_0=  a+b+c+d+e+f+g+h+i+2j+2k+2l\eq$$

Now express the elementary
couplings in terms of dummy variables $\{L_1,L_2, M_1, M_2, R_1, R_2, P, Q\}$
whose exponent are the  BZ basis data, respectively $\{\la_1, \la_2, \mu_1,
\mu_2, r_1,r_2,p, q\}$:
$$\eqalignT{
&A_1= M_1\quad 
&A_2=L_1\quad 
&A_3= L_1M_1PQ\cr
&B_1=M_2\quad 
&B_2= L_2\quad 
&B_3= L_2M_2R_1^2R_2^2\cr
&C_1= L_2M_1Q \quad
&C_2= L_1M_2R_2^2P \quad
&C_3= L_1M_1P\cr
&D_1=L_1^2M_2R_2^2P^2\quad
&D_2= L_2M_1^2R_1^2 \quad
&D_3= L_2M_2R_2^2 \cr}\eq$$
Next, consider each term of the generating function (\genfusp) and solve for $k_0$
in terms of the basis variables. Surprisingly
there are only four different formulas for 
$k_0$.  The expressions corresponding to the different terms of (\genfusp) are:
$$\eqalignD{
& {\rm terms}~1,5,6: \quad  &k_0= \la_1+\la_2+\mu_1+\mu_2-p-q-r_1/2\cr
& {\rm terms}~3,8: \quad  &k_0= \la_1+\mu_1+\mu_2-p\cr
& {\rm term~}~7: \quad  &k_0= \la_1+\la_2+\mu_2+r_1/2-r_2\cr
& {\rm terms}~2,4: \quad  &k_0= \la_1+\la_2+\mu_2-r_2/2\cr}\eqlabel\tresp$$
Therefore, the threshold formula is the maximum value of these four values or
equivalently
$$k_0= \la_1+\la_2+\mu_1+\mu_2- {\rm min} (p+q+r_1/2,\,
\la_2+p,\, \mu_1 - r_1/2+r_2,\, \mu_1+r_2/2)\eq$$
Notice that by rewriting $k\geq k_0$, we recover from (\tresp) the 4 inequalities 
(\trespbas).

The system of inequalities (\bzspq) can be transformed into a system of
equations by setting 
$r_1/2=s_1$ and $r_2/2=s_2$ and introducing the integers $a_i$
(cf. section 7.5 of [\BCM]):
$$\eqalignD{
&\lambda_1  =p+a_1 \qquad &\nu_2  =a_4+a_8 \cr
&\lambda_2  =s_1+a_2 \qquad &a_2+p =a_3+q \cr
&\mu_1  =q+a_5  \qquad &        a_3+s_1  =a_4+s_2 \cr
&\mu_2  =s_2+a_8 \qquad &a_5+2s_{2}  =a_6+2s_{1} \cr
&\nu_1  =a_1+a_7 \qquad & a_6+q  =a_7+p \cr}   \eq$$
As shown in [\BCM], this leads to the following diamond-type graphical representation
of the tensor product:

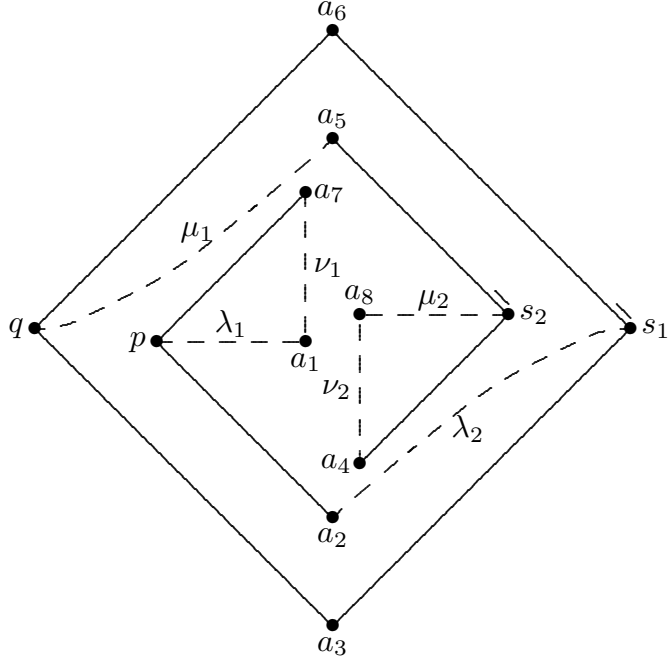
\input pictex

\def\bpc{\beginpicture}

\def\epc{\endpicture}

\def\figureF{
\bpc
        \setcoordinatesystem units <0.6mm,0.6mm>
        \setplotarea x from -55 to 55, y from -55 to 55
        \setlinear
        \plot 0 55 -55 0 0 -55 55 0 0 55 /
        \plot 5 -25 32.5 2.5  0 35 /
\plot -5 25 -32.5 -2.5  0 -35 /
\plot 32.5 4.5 30 7 /
        \plot 55 2 52.5 4.5 /
        \multiput {$\bullet$} at 0 55 0 -55 55 0 -55 0 0 35 0 -35 32.5 2.5 -32.5 -2.5 5 -25 -5 25 5 2.5 -5 -2.5
/
        \put {$s_1$} [l] at     57 0
        \put {$s_2$} [l] at 34.5 2.5
        \put {$q$} [r] at -57 0
        \put {$p$} [r] at -34.5 -2.5
        \put {$\lambda_1$} [b] at -18.75 -1.5
        \put {$\mu_2$} [b] at 18.75 3.5
        \put {$\nu_1$} [l] at -3.5 11.25
        \put {$\nu_2$} [r] at 3.5 -11.25
        \put {$\mu_1$} [b] at -25 16
        \put {$\lambda_2$} [t] at 25 -16
        \put {$a_6$} [b] at 0 57
        \put {$a_5$} [b] at 0 37
        \put {$a_2$} [t] at 0 -37
        \put {$a_3$} [t] at 0 -57
        \put {$a_4$} [r] at 3.5 -25
        \put {$a_7$} [l] at -3.5 25
        \put {$a_8$} [b] at 5 4.5
        \put {$a_1$} [t] at -5 -4.5
        \setdashes
        \plot 32.5 2.5 5 2.5 5 -25 /
        \plot -32.5 -2.5 -5 -2.5 -5 25 /
        \setquadratic
        \plot 0 -35 5 -30.5 15 -22 20 -18 25 -14 30 -10.5 35 -7.5 40 -5 45 -2.5 50 -0.5 55 0
/
        \plot 0 35 -5 30.5 -15 22 -20 18 -25 14 -30 10.5 -35 7.5 -40 5 -45 2.5 -50 0.5 -55 0
/
\epc}

\vskip1cm \centerline{ \figureF} \vskip1cm

Dotted lines relate those two points that compose the label indicated beside it and
opposite continuous lines are constrained to be equal, with the length of a line
being defined as the sum of its extremal points except for the lines delimited by
the points
$(a_6, s_1)$ and $(a_5,s_2)$ where the point $s_i$ is
counted twice (the little bar besides $s_1$ and $s_2$ being a
reminder of this particularity).  For those lines,  the
constraint reads
$a_6+2s_1=a_5+2s_2$.  Given a triple
$sp(4)$ product, the number of such diamonds that 
can be drawn with only non-negative
entries gives its multiplicity.

In terms of these data, the expression for the threshold level (\tresp) look somewhat
more symmetrical: the four expressions in (\tresp) correspond respectively  to the
following terms:
$$k_0 = a_1+a_8+{\rm max}\{ a_4+a_7+ s_1,
 a_5+q+s_2, a_4+q+s_1, a_4+q+s_2\}\eq$$


\newsec{The $\su(4)$ generating function}

Written directly in terms of LR tableaux, the $su(4)$ elementary solutions are:
$$\eqalign{
A_1 &=\ST{\STrow{\b1}\STrow{\b2}\STrow{\b3} }, \quad  
 A_2=\ST{\STrow{\bv}\STrow{\bv}\STrow{\bv}\STrow{\b1}}, \quad
 A_3=\ST{\STrow{\bv}},\quad
B_1 =\ST{\STrow{\b1}\STrow{\b2} }, \quad  
 B_2=\ST{\STrow{\bv}\STrow{\bv}\STrow{\b1}\STrow{\b2}}, \quad
 B_3=\ST{\STrow{\bv}\STrow{\bv}},\cr
C_1 &=\ST{\STrow{\b1} }, \quad  
 C_2=\ST{\STrow{\bv}\STrow{\b1}\STrow{\b2}\STrow{\b3}}, \quad
 C_3=\ST{\STrow{\bv}\STrow{\bv}\STrow{\bv}},\quad
D'_1 =\ST{\STrow{\bv}\STrow{\bv}\STrow{\b1} }, \quad  
 D'_2=\ST{\STrow{\bv}\STrow{\b1}}, \quad
 D'_3=\ST{\STrow{\bv}\STrow{\b1}\STrow{\b2}},\cr}\eqlabel\elequ$$
and
$$
D_1 =\ST{\STrow{\bv\b1}\STrow{\bv}\STrow{\b2}\STrow{\b3} }, \; 
D_2 =\ST{\STrow{\bv\b1}\STrow{\bv\b2}\STrow{\bv}\STrow{\b3} }, \; 
D_3 =\ST{\STrow{\bv\b1}\STrow{\bv}\STrow{\bv}\STrow{\b2} },\; 
E_1 =\ST{\STrow{\bv\bv}\STrow{\bv\b1}\STrow{\bv}\STrow{\b2} }, \; 
E_2 =\ST{\STrow{\bv\b1}\STrow{\bv}\STrow{\b2} }, \,
E_3 =\ST{\STrow{\bv\b1}\STrow{\bv\b2}\STrow{\b1}\STrow{\b3} }
\eqlabel\elequu$$ 
The
relations are [\ref{R.T Sharp and D. Lee, Revista Mexicana de Fisica {\bf 20}(1971)
203.}\refname\SL,\ref{M.Couture,
C.J.Cummins and R.T.Sharp, J.Phys {\bf A23} (1990) 1929.}\refname\CCS]:
$$ \eqalignT{ &D_j^{'}  D_ k = C_i E_i \qquad &D_j D_k^{'}  = B_i C_j C_k  \qquad
&E_i E_j  = B_k D_k D_k^{'} \cr
&D_i E_i  = C_j B_k D_k \qquad
&D_i^{'}E_i   = B_j D_j^{'} C_k \qquad &\quad {} \cr}
\eqlabel\mmmhb$$
with $i,j,k$ a cyclic permutation of $1,2,3$. 

Consider now the construction of the set of fusion elementary couplings using
outer-automorphism completeness.  Start with
$A_1:(0,0,0)\otimes(0,0,1)\supset(0,0,1)$, this has threshold level 1.  Acting on it
with 
$$(A,A') = (a^n,a^m),\qquad  n,m=0,1,2,3\eq$$ where $$
a[\la_0,\la_1\,\la_2,\la_3]= [\la_3,\la_0\,\la_1,\la_2]\eq$$ we generate the affine
extension of the whole set $A_i,B_i,C_i,D_i, D'_i$, which thus all have threshold
level 1, together with the scalar coupling
$\E_0= [1,0,0,0]\times [1,0,0,0]\supset[1,0,0,0]$. Finally, the affine extension of
$E_1$ arises first at level 2: $[0,1,0,1]\times[1,0,1,0]\supset[1,0,1,0]$.  The
three weights in this coupling are fixed under the action of $A=a^2$. Hence, we need
only to consider 
$$(A,A')= \{(1,1), (a,a),(a,1),(1,a)\}\eq$$ and this leads respectively to
$\E_1, \E_2, \E_3$, which all have $k_0=2$ and a new elementary coupling
$\F=[0,1,0,1]\times[0,1,0,1]\supset[0,1,0,1]$ (first discovered in
[\BKMW]). Notice that at the level of tensor-products, $\F$ is a composite product
$C_1C_2C_3$.  But if it were still composite for fusions, it would necessarily have
level 3 since $k_0(C_1C_2C_3)=3$. This is the reason why $\F$ must be regarded
as a new elementary coupling.

The whole set of fusion elementary couplings
is:
$$\eqalignD{
&\A_1=[1,0,0,0]\times[0,0,0,1]\supset[0,0,0,1]\qquad
&\D_1'=[0,0,1,0]\times[0,1,0,0]\supset[0,0,0,1]\cr
&\A_2=[0,0,0,1]\times[0,1,0,0]\supset[1,0,0,0]\qquad
&\D_2'=[0,1,0,0]\times[0,1,0,0]\supset[0,0,1,0]\cr
&\A_3=[0,1,0,0]\times[1,0,0,0]\supset[0,1,0,0] \qquad
&\D_3'=[0,1,0,0]\times[0,0,1,0]\supset[0,0,0,1] \cr 
&\B_1=[0,0,0,0]\times[0,0,1,0]\supset[0,0,1,0]\qquad
&\D_1=[0,0,1,0]\times[0,0,0,1]\supset[0,1,0,0]\cr 
&\B_2=[0,0,1,0]\times[0,0,1,0]\supset[1,0,0,0]\qquad 
&\D_2=[0,0,0,1]\times[0,0,0,1]\supset[0,0,1,0]\cr 
&\B_3=[0,0,1,0]\times[1,0,0,0]\supset[0,0,1,0]\qquad
&\D_3=[0,0,0,1]\times[0,0,1,0]\supset[0,1,0,0] \cr 
&\C_1=[1,0,0,0]\times[0,1,0,0]\supset[0,1,0,0]\qquad 
&\E_1=[0,1,0,1]\times[1,0,1,0]\supset[1,0,1,0]\cr 
&\C_2=[0,1,0,0]\times[0,0,0,1]\supset[1,0,0,0]  \qquad
&\E_2=[1,0,1,0]\times[1,0,1,0]\supset[0,1,0,1]\cr  
&\C_3=[0,0,0,1]\times[1,0,0,0]\supset[0,0,0,1]\qquad
&\E_3=[1,0,1,0]\times[0,1,0,1]\supset[1,0,1,0]\cr}\eq$$
together with two couplings that have no elementary finite relative:
$$ \E_0=[1,0,0,0]\times[1,0,0,0]\supset[1,0,0,0]\qquad
\F=[0,1,0,1]\times[0,1,0,1]\supset[0,1,0,1]\eq$$

 The tensor-product
 relations are modified by the appropriate insertions of $d$ or $\E_0$ factors in
order to put them at the same threshold level:
$$ \eqalignT{ &\E_0\D_j^{'}  \D_ k = \C_i \E_i \qquad &\E_0\D_j \D_k^{'}  = \B_i\C_j
\C_k 
\qquad &\E_i \E_j  = \E_0\B_k \D_k \D_k^{'} \cr
&\D_i \E_i  = \C_j \B_k \D_k \qquad
&\D_i^{'}\E_i   = \B_j \D_j^{'} \C_k \qquad &\E_0\F= \C_1\C_2\C_3 {} \cr}
\eqlabel\mmmhb$$
with $i,j,k$ a cyclic permutation of $1,2,3$.

To get the $\su(4)$
 basis, we first write down the $V$ matrix, whose columns are
the vectorial transcription of the fusion elementary couplings written in terms
of the grading variables.  The column ordering corresponds to 
$\E_0,
\A_i,
\B_i,
\C_i,
\D'_i,
\D_i,
\E_i, \F$ with $i=1,2,3$.   The rows are labelled by
the LR variables $(k, \la_1, \la_2, \la_3, n_{11}, n_{12},n_{13}, n_{14},n_{22},
n_{23},n_{24}, n_{33},n_{34})$.  The matrix $V$ is thus
$$V=\pmatrix {
1&1&1&1&1&1&1&1&1&1&1&1&1&1&1&1&2&2&2&2&\cr
0&0&0&1&0&0&0&0&1&0&0&1&1&0&0&0&1&0&0&1&\cr
0&0&0&0&0&1&1&0&0&0&1&0&0&1&0&0&0&1&1&0&\cr
0&0&1&0&0&0&0&0&0&1&0&0&0&0&1&1&1&0&0&1&\cr
0&1&0&0&1&0&0&1&0&0&0&0&0&1&1&1&0&1&1&1&\cr
0&0&0&0&0&0&0&0&1&0&0&1&1&0&0&0&1&0&0&1&\cr
0&0&0&0&0&1&0&0&0&0&1&0&0&0&0&0&0&0&1&0&\cr
0&0&1&0&0&0&0&0&0&0&0&0&0&0&0&0&0&0&0&0&\cr
0&1&0&0&1&0&0&0&0&0&0&0&0&0&1&0&0&0&1&0&\cr
0&0&0&0&0&0&0&0&1&0&0&0&1&1&0&0&0&1&0&1&\cr
0&0&0&0&0&1&0&0&0&0&0&0&0&0&0&1&1&0&0&0&\cr
0&1&0&0&0&0&0&0&0&0&0&0&0&0&0&0&0&0&0&0&\cr
0&0&0&0&0&0&0&0&1&0&0&0&0&1&1&0&0&0&1&1&\cr}\eq$$
With $u=(u_0, \cdots, u_{12})$, the equations $u^\top\,  V\geq 0$ can be
transformed into equalities by introducing the variables $\a_i$:
$$\eqalignD{
&u_0= \a_0\qquad &u_0+u_2+u_6= \a_{10}\cr
&u_0+u_4+u_8+u_{11}= \a_1\qquad &u_0+u_1+u_5= \a_{11}\cr
&u_0+u_3+u_7= \a_2\qquad  &u_0+u_1+u_5+u_9= \a_{12}\cr
&u_0+u_1= \a_3\qquad  &u_0+u_2+u_4+u_9+u_{12}= \a_{13}\cr
&u_0+u_4+u_8= \a_4\qquad  &u_0+u_3+u_4+u_8+u_{12}= \a_{14}\cr
&u_0+u_2+u_6+u_{10}= \a_5\qquad  &u_0+u_3+u_4+u_{10}= \a_{15}\cr
&u_0+u_2= \a_6\qquad  &2u_0+u_1+u_3+u_5+u_{10}= \a_{16}\cr
&u_0+u_4= \a_7\qquad &2u_0+u_2+u_4+u_9= \a_{17}\cr
&u_0+u_1+u_5+u_9+u_{12}= \a_8\qquad  &2u_0+u_2+u_4+u_6+u_8+u_{12}= \a_{18}\cr
&u_0+u_3= \a_9\qquad 
&2u_0+u_1+u_3+u_4+u_5+u_9+u_{12}= \a_{19} \cr}\eq$$

We have 13 free variables; let us choose
them to be the $\a_i$ for $i=0, \cdots, 12$.  Solving for the dependent variables
leads to
$$\eqalignT{
&u_0= \a_0\qquad       &u_5= -\a_3+\a_{11}\qquad   &u_{10}= \a_{5}-\a_{10}    \cr
&u_1= -\a_0+\a_3\qquad &u_6= -\a_6+\a_{10}\qquad   &u_{11}= \a_{1}-\a_{4} \cr
&u_2= -\a_0+\a_6\qquad &u_7= \a_2-\a_9 \qquad   &u_{12}= \a_{8}-\a_{12} \cr
&u_3= -\a_0+\a_9\qquad &u_8= \a_4-\a_7  \qquad \cr
&u_4=-\a_0+\a_7\qquad  &u_9= -\a_{11}+\a_{12} \qquad \cr}\eq$$
together with
$$\eqalignD{& \a_{13}= -\a_0+\a_6+\a_7+\a_8-\a_{11}\qquad & 
\a_{17}=\a_6+\a_7-\a_{11}+\a_{12}\cr
&\a_{14}= -\a_0+\a_4+\a_8+\a_9-\a_{12}\qquad & 
\a_{18}=\a_4+\a_8+\a_{10}-\a_{12}\cr 
& \a_{15}= -\a_0+\a_5+\a_7+\a_9-\a_{10}\qquad
&\a_{19}= -\a_0+\a_7+\a_8+\a_{9}\cr 
& \a_{16}= \a_5+\a_9-\a_{10}+\a_{11}\qquad
&\cr}\eqlabel\ssff$$

Now, by setting successively $\a_i=1$ for $i=0, \cdots, 12$ and the others equal to
0, we generate the following set of basis vectors:
$$\eqalignD{
&\Ec_0=
dN_{11}^{-1}L_1^{-1}L_2^{-1}L_3^{-1}\Ac_0\Ac_{13}^{-1}\Ac_{14}^{-1}
\Ac_{15}^{-1}\Ac_{19}^{-1}\qquad &\Ec_6=  N_{13}^{-1}L_2 \Ac_6\Ac_{13}\Ac_{17}\cr
&\Ec_1= N_{33}\Ac_1\qquad &\Ec_7= N_{11}
N_{22}^{-1}L_2\Ac_7\Ac_{13}\Ac_{15}\Ac_{17}\Ac_{19}\cr &\Ec_2= N_{14}\Ac_2\qquad
&\Ec_8=  N_{34}\Ac_8\Ac_{13}\Ac_{14}\Ac_{18}\Ac_{19}\cr &\Ec_3=
N_{12}^{-1}L_1\Ac_3\qquad &\Ec_{9}= 
N_{14}^{-1}L_3\Ac_9\Ac_{14}\Ac_{15}\Ac_{16}\Ac_{19}\cr  &\Ec_4=
N_{22}^{-1}N_{33}\Ac_4\Ac_{14}\Ac_{18}\qquad &\Ec_{10}= 
N_{13}N_{24}^{-1}\Ac_{10}\Ac_{15}^{-1}\Ac_{16}^{-1}\Ac_{18}\cr
 &\Ec_5= N_{24}\Ac_{5}\Ac_{15}\Ac_{16}\qquad &\Ec_{11}= N_{12}
N_{23}^{-1}\Ac_{11}\Ac_{13}^{-1}\Ac_{16}\Ac_{17}^{-1}\cr
& &\Ec_{12}= N_{23}
N_{34}^{-1}\Ac_{12}\Ac_{14}^{-1}\Ac_{17}\Ac_{18}^{-1}\cr  }
\eqlabel\rtrt$$

We must now look for those combinations that contain only non-negative powers of the
$\Ac_i$. Since each $\Ec_i$ contains at least one positive power of $\Ac_i$,
these must be obtained from positive combinations of the $\Ec_j$.  To find
them, it is convenient to proceed as in the analysis of $\sp(4)$.
Denote by $a_i$ the number of ${\cal A}_i$ factors in a general term $\prod {\cal
E}_i^{\e'_i}$ of the free expansion of the $\Ec_i$ in non-negative powers we get
(equivalently, we can read off the  
${\cal A}_i$ from  (\rtrt)):
$$\eqalignD{
& a_{13}= -\e'_0+\e'_6+\e'_7+\e'_8-\e'_{11}\qquad 
& a_{17}=\e'_6+\e'_7-\e'_{11}+\e'_{12}\cr 
& a_{14}=-\e'_0+\e'_4+\e'_8+\e'_{9}-\e'_{12}\qquad  
& a_{18}=\e'_4+\e'_8+\e'_{10}-\e'_{12}\cr 
& a_{15}=-\e'_0+\e'_5+\e'_7+\e'_{9}-\e'_{10}\qquad & a_{19}=
-\e'_0+\e'_7+\e'_8+\e'_{9}\cr
& a_{16}=\e'_5+\e'_{9}-\e'_{10}+\e'_{11}\qquad ~\cr}\eq$$ 
These relations are to be
compared with (\ssff).  We thus look for elementary solutions of the system
$a_i\geq0$ for  
$\e'_i$ non-negative integers. 

 The full list of composites
that involve
$\Ec_0$ -- these are those that generate
$k$-dependent constraints 
 --  is
$$\eqalignS{ &\Ec_0\Ec_4\Ec_7,\quad  &\Ec_0\Ec_8\Ec_{9},\quad  
&\Ec_0\Ec_6\Ec_{9},\quad  & \Ec_0\Ec_9\Ec_8, \quad   &\Ec_0\Ec_7\Ec_{9},\quad  
&\Ec_0\Ec_5\Ec_{8},\cr &\Ec_0\Ec_7\Ec_8\Ec_{11},\quad 
&\Ec_0\Ec_7\Ec_{9}\Ec_{12},\quad &\Ec_0\Ec_7\Ec_{9}\Ec_{10},\quad 
&\Ec_0\Ec_5\Ec_{8}\Ec_{12}\Ec_4\quad &\Ec_0^2\Ec_7\Ec_8\Ec_9\cr}\eq$$

The constraints are  ($\Ec_i$ specifies the vector $e_j$ such that the inequality is
$e_j\, x\geq 0)$:
$$\eqalignD{ 
&\Ec_0\Ec_4\Ec_7:\quad  &k \geq \la_1+\la_2+\la_3+n_{33}\cr
& \Ec_0\Ec_8\Ec_{9}:\quad & k \geq \la_1+\la_2+n_{11}+n_{14}-n_{34}\cr
& \Ec_0\Ec_6\Ec_{9}:\quad &  k \geq \la_1+n_{11}+n_{13}+n_{14}\cr
&\Ec_0\Ec_7\Ec_8: \quad &  k \geq \la_1+\la_2+\la_3+n_{22}-n_{34}\cr
&\Ec_0\Ec_7\Ec_{9}:\quad & k \geq \la_1+\la_2+n_{14}+n_{22}\cr
&\Ec_0\Ec_5\Ec_{8}: \quad & k \geq \la_1+\la_2+\la_3+n_{11}-n_{24}-n_{34}\cr
&\Ec_0\Ec_7\Ec_8\Ec_{11}:\quad 
& k \geq \la_1+\la_2+\la_3-n_{12}+n_{22}+n_{23}-n_{34}\cr
&\Ec_0\Ec_7\Ec_{9}\Ec_{10}:\quad  &k \geq
\la_1+\la_2+n_{14}-n_{13}+n_{22}+n_{24}\cr
&\Ec_0\Ec_8\Ec_{9}\Ec_{12}: \quad &k \geq
\la_1+\la_2+n_{11}+n_{14}-n_{23}\cr
&\Ec_0^2\Ec_7\Ec_8\Ec_9 : \quad & 2 k \geq 2\la_1+2\la_2+\la_3
+n_{14}+n_{22}+n_{11}-n_{34}\cr }\eqlabel\kokir$$
When these inequalities are re-expressed in terms of BZ triangle data, they
reproduce the threshold formula presented in [\BKMW]. 

The $\Ec_0$-independent
elementary solutions, namely
$$\eqalign{&\Ec_1,\quad
\Ec_2,\quad\Ec_3,\quad\Ec_4,\quad\Ec_5,\quad\Ec_6,\quad\Ec_7,\quad
\Ec_8,\quad\Ec_{9}\cr
&\Ec_6\Ec_{11},\quad\Ec_4\Ec_{12}
,\quad\Ec_7\Ec_{11},\quad\Ec_8\Ec_{12},\quad\Ec_9\Ec_{10},\quad \Ec_5\Ec_{10},\cr
&\Ec_8\Ec_{11}\Ec_{12},\quad\Ec_9\Ec_{10}\Ec_{12},\quad\Ec_7\Ec_{10}\Ec_{11}\cr}\eq$$
yield the standard LR inequalities:
$$\eqalignD{
&\la_1 \geq n_{12}\qquad &n_{11} \geq n_{22}\cr
&\la_2 \geq n_{13}\qquad & n_{11}+n_{12} \geq n_{22}+ n_{23}\cr
&\la_2+n_{12} \geq n_{13}+n_{23}\qquad &n_{11}+n_{12}+n_{13} \geq n_{22}+
n_{23}+n_{24}\cr
 &\la_3 \geq n_{14}\qquad &n_{22} \geq n_{33}\cr
&\la_3+n_{13} \geq n_{14}+n_{24}\qquad &n_{22}+n_{23} \geq n_{33}+ n_{34}\cr
&\la_3+n_{13}+n_{23} \geq n_{14}+n_{24}+  n_{34}\qquad ~\cr
}\eqlabel\inequatre$$
and  $n_{ij}\geq 0$, except for $n_{11}\ge 0 $ and
$n_{22}\geq 0$ which are implied  by the above equations.

As in the $\sp(4)$ case, we can check that the relations (\ssff) code the `basic
linear relations' of the model.  Indeed, the 7 relations read from (\ssff) are
$$\eqalignD{
&\E_0\D_1\D_2'= \B_3\C_1\C_2\qquad &\E_1\D_1'= \B_2\C_3\D_2'\cr
&\E_0\D_2\D_3'= \B_1\C_2\C_3\qquad &\E_2\D_2'= \B_3\C_1\D_3'\cr
 &\E_0\D_3\D_1'= \B_2\C_1\C_3\qquad &\E_3\D_3'= \B_1\C_3\D_1'\cr
&\E_0\F= \C_1\C_2\C_3\qquad ~&\cr}\eq$$
and these are the generators of all the $\su(4)$ linear relations.


In order to construct the $\su(4)$ generating function, we must choose a term
ordering.  We fix the ordering as follows:
$$\eqalign{
& \{L_1,L_2,L_3,N_{11},N_{12},N_{13},N_{14},N_{22},N_{23},N_{24},N_{33},
N_{34},d,
\cr 
& \E_1,\E_2,\E_3,\B_1,\B_2,\B_3,\C_1,\C_2,\C_3,\A_1,\A_2,\A_3,\D_1,
\D_2,\D_3,\D_1^{'},\D_2^{'},\D_3^{'},\E_0,\F \}\cr}\eq$$
 Grobner basis methods yield  the forbidden
products:
$$\eqalign{ \{ & \E_i \E_j,\, \D_i^{'}
\E_i,\, \D_i \E_i,\, \C_i \E_i, \,\B_i \C_j \C_k,\,\B_2 \C_3 \D_1 \D_2^{'},\,\B_1
\C_3
\D_1
\D_2^{'},\,\B_1 \C_2 \D_3 \D_1^{'}, \cr
& \E_0 \B_1 \D_1 \D_3 \D_1^{'}  \D_2^{'},\,\F \E_i,
\F \B_i, \,\C_1 \C_2 \C_3, \,\F \C_1 \C_2 \C_3\} \cr}. \eq$$ 
The different terms of the generating function are fully specified by their
denominator (the numerators are introduced to avoid over-counting):
$$\eqalignD{
& \E_0 \A_1 \A_2 \A_3 \D_1 \D_2 \D_3 \D_1^{'} \D_2^{'} \D_3^{'} \C_1 \C_2 \F, \quad
&\E_0 \A_1 \A_2 \A_3 \B_2 \B_3 \C_2 \C_3 \D_2 \D_3 \D_2^{'} \D_3^{'} \E_1 \F,
\cr  & \E_0 \A_1 \A_2 \A_3 \B_1 \B_2 \B_3 \C_2 \D_1 \D_2 \D_1^{'} \D_2^{'} \E_3 \F,
\quad
&\E_0 \A_1 \A_2 \A_3 \B_1 \B_2 \B_3 \C_1 \D_1 \D_2 \D_1^{'} \D_2^{'} \E_3 \F, \cr
&\E_0 \A_1 \A_2 \A_3 \B_2 \C_1 \C_2 \D_1 \D_2 \D_3 \D_1^{'} \D_2^{'} \D_3^{'},\quad
&\E_0 \A_1 \A_2 \A_3 \B_1 \B_2 \C_1 \C_2 \D_1 \D_2 \D_1^{'} \D_2^{'} \D_3^{'},\cr
& \E_0 \A_1 \A_2 \A_3 \B_1 \B_2 \C_1 \C_2 \D_1 \D_2 \D_3 \D_2^{'} \D_3^{'},\quad
&\E_0 \A_1 \A_2 \A_3 \B_3 \C_1 \C_3 \D_1 \D_2 \D_3 \D_1^{'} \D_2^{'} \D_3^{'}, \cr
& \E_0 \A_1 \A_2 \A_3 \B_1 \B_3 \C_1 \C_3 \D_2 \D_3 \D_1^{'} \D_2^{'} \D_3^{'}, \quad
&\E_0 \A_1 \A_2 \A_3 \B_1 \B_3 \C_1 \C_3 \D_1 \D_2 \D_1^{'} \D_2^{'} \D_3^{'},\cr
& \E_0 \A_1 \A_2 \A_3 \B_1 \B_3 \C_1 \C_3 \D_1 \D_2 \D_3  \D_2^{'} \D_3^{'},\quad
& \E_0 \A_1 \A_2 \A_3 \B_1 \B_3 \C_1 \C_3 \D_1 \D_2 \D_3 \D_1^{'} \D_3^{'}, \cr 
& \A_1 \A_2 \A_3 \B_1 \B_3 \C_1 \C_3 \D_1 \D_2 \D_3 \D_1^{'} \D_2^{'} \D_3^{'},\quad
&\A_1 \A_2 \A_3 \B_1 \B_2 \B_3 \C_1 \D_1 \D_2 \D_3 \D_1^{'} \D_2^{'} \D_3^{'},\cr
& \E_0 \A_1 \A_2 \A_3 \B_2 \B_3 \C_1 \D_1 \D_2 \D_3 \D_1^{'} \D_2^{'} \D_3^{'},\quad
& \E_0 \A_1 \A_2 \A_3 \B_1 \B_2 \B_3 \C_1 \D_2 \D_3 \D_1^{'} \D_2^{'} \D_3^{'}, \cr
& \E_0 \A_1 \A_2 \A_3 \B_1 \B_2 \B_3 \C_1 \D_1 \D_2 \D_1^{'} \D_2^{'} \D_3^{'},\quad
& \E_0 \A_1 \A_2 \A_3 \B_1 \B_2 \B_3 \C_1 \D_1 \D_2 \D_3 \D_2^{'} \D_3^{'}, \cr
& \E_0 \A_1 \A_2 \A_3 \B_1 \B_2 \B_3 \C_1 \D_1 \D_2 \D_3 \D_1^{'} \D_3^{'},\quad
\cr}\eq$$

Finally, note that the expression of threshold levels in terms of tableau data,
that is, the analogue of the
$\su(2,3)$ formulas written previously is clearly
$$k_0 = \# {\rm columns}- {\rm max} \{\#  D_1+\#  D_2+\#  D_3+\#  C_1C_2C_3\}\eq$$ 
since the $D_i's$ have level 1 but two columns and $C_1C_2C_3$ has level 2 and three
columns (corresponding to the $\F$ fusion coupling.)

\newsec{Conclusion and open problems}

We have obtained the fusion generating
function for
$\su(3,4)$ and $\sp(4)$ using the  conjectural existence of a fusion basis. In
the $\su(3)$ case a first-principle  derivation (presented in Appendix A)
provides an independent proof of the results, thus a partial confirmation of
the conjectures and the correctness of the underlying fusion basis.
Moreover, different tests of the
$\sp(4)$ and
$\su(4)$ generating functions, presented in Appendix A, also support our conjectures
and the fusion basis constructions. 
{\it En passant}, we point out that the search for the complete $\su(N)$
level-rank symmetric function introduced in Appendix A  is a quest that
deserves further studies.

Although the theme of this paper is the construction of fusion generating
functions, our most important result is the unravelling of the fusion basis
concept, for which we have provided concrete examples.  The
main open problem is to find a fundamental and Lie algebraic way of deriving the
fusion basis (analogous  to the Berenstein-Zelevinsky conjectures [\BZin]).  We
observe that the number of
$k$-type inequalities increases rather quickly with the rank of the algebra: 1
for $\su(2)$, 3 for
$\su(3)$, 4 for $\sp(4)$ and 10 for
$\su(4)$.
More specifically  we would  like to find arguments to
justify the homogeneity property (on the other hand, the linearity appears to be a
generic property, a direct consequence of the Kac-Walton algorithm). 


With regard to the automorphism completeness
conjecture we note that
for
simplicity (and because the discussion is to a large extent devoted to $\su(N)$ for
which the outer-automorphism group is rather large) we have focused on the
outer-automorphism group as the essential symmetry.  It is 
natural to extend the conjecture to the full symmetry group of fusion
coefficients. 
However,  we should stress is that the outer-automorphism conjecture is just a
convenient tool. If the
conjecture (or its natural extension to the full fusion symmetry group) turns out to
be wrong, there are other avenues that could yield  the complete set of fusion
elementary couplings.


In the present work, the only information on fusion data that has been 
extracted, out of the fusion basis or the fusion generating function, 
is 
the expression for the threshold level in terms 
of the basis variables. But
there are certainly more data that can be lifted.  
For example, given a triple
product with multiplicity $m$, to which there correspond $m$ values of the threshold
levels, we could ask for the expression, 
in terms of the  Dynkin labels, of the 
minimum and maximum values of
$k_0$.  It is easy to write down some explicit
expressions for particular fusion coefficients.

The reformulation of the problem of computing fusion rules
in terms of a fusion basis solves, in principle, the quest for
a combinatorial method since it  reduces a fusion computation to solving
inequalities. But we expect that we have not found an optimal solution to the quest
for an efficient combinatorial description. 


\smallskip
\centerline{\bf Acknowledgement}
L.B. would like to thank P. Dargis for his crucial, albeit involuntary, r\^ole in
bringing Farkas' lemma to his attention.
\appendix{A}{Independent verifications of the fusion generating functions}

 The $\su(3)$
fusion generating function is not presented here for the first time; it appeared
originally in [\CMW]. A sketch of its proof was presented in [\ref{C.J.
Cummins,
J. Phys. A 24
(1991) 391.}\refname\Ca]
without  details.  
In this section we present a complete proof of the $\su(3)$ generating function for
fusion rules; in addition, we describe some independent checks
confirming the validity of  the
$\su(4)$ and
$\sp(4)$ fusion generating functions given in sections 6.3 and 7.3. The first check
that we present uses Giambelli-type  formulas.  These can be viewed as equalities
of  corresponding expressions in the character rings. Since the 
$\su(n)$ and $\sp(n)$ fusion rings are quotients of
the classical character rings 
(see [\ref{J. Fuchs,
Fortschr. Phys. 42 (1994) 1. }\refname\FUCH]
and references therein),
these formulas
continue to hold for fusion products.  For $\su(4)$, we present another
non-trivial check based on a level-rank duality argument.

\def\OMun{{\displaystyle\Om_{\geq}^{M_1}}}

\subsec{Determinantal formula and the `composition' method: deriving the
$su(3)$ generating function for tensor products}

The Giambelli formula, or more generally, determinantal
formulae which give expressions
for group characters as determinants, provide  another method for 
calculating fusion generating functions in terms 
of simpler generating functions. This uses the technique of `composition' of
generating functions described previously in section 2.3 of [\BCM].

The $su(3)$ Giambelli formula  expresses a
general
representation in terms of a
difference of products of representations with a
single non-zero
Dynkin label,
i.e.,
$$(\la_1,\la_2) = (\la_1+\la_2,0)\otimes
(\la_2,0)-
 (\la_1+\la_2+1,0)\otimes (\la_2-1,0)\eqlabel\giam$$
This can be rewritten in determinantal form as follows
$$
(\lambda_1,\lambda_2) =\det
\pmatrix{ (\lambda_1+\la_2,0) & (\lambda_2-1,0)\cr 
(\lambda_1+\la_2+1,0) &(\lambda_2,0) \cr }
\eq$$

Consider first the generating function
$
G_1(L_1,L_2,M_1,R_1,R_2)$
which is the generating function for products
of the form:
$(\la_1, \la_2)\otimes (\mu_1,0)$.  Its explicit form is
$$\eqalign{
G_1 =
{1\over(1-L_1N_1)(1-L_2N_2)(1-L_2M_1)(1-M_1N_1)
(1-L_1M_1N_2)}\cr}
\eq$$
It is obtained by setting $M_2=0$ in the complete tensor-product generating
function (cf. section 2.5 in [\BCM]).  Our point here is not to re-derive $G_1$
from first principles but simply to show how we can  reconstruct the complete
generating function out of the partial information contained in $G_1$.  In the
fusion case, we will indicate how the analogue of $G_1$ can be obtained,
preventing the argument from being circular.

{}From two copies of $G_1$ we form the composite generating function $G_2$:
$$
G_2(L_1,L_2,M_1,M_2,N_1,N_2)=
\OR\, 
G_1(L_1,L_2,M_1,R_1,R_2)G_1(R_1^{-1},R_2^{-1},M_2,N_1,N_2)\eq
$$
which is the generating function for
products of the form
$$(\la_1, \la_2)\otimes (\mu_1,0)\otimes
(\mu_2,0)\eq$$ 
Note that the generating function for
products $$(\la_1, \la_2)\otimes (\mu_1+1,0)\otimes
(\mu_2-1,0)\eq$$ is $ M_2M_1^{-1}G_2$ and so, by (\giam),
 the generating function for products
$(\la_1, \la_2)\otimes (\mu_1,\mu_2)$ is:
$$
G_3= \OMun\, ( G_2 - M_2M_1^{-1}G_2)\eq
$$
The coefficient of $M_1^{\mu_1}M_2^{\mu_2}$ is the
multiplicity of the representation with Dynkin
labels
$(\mu_1-\mu_2,\mu_2)$ in the product $$(\la_1,
\la_2)\otimes
\left[ (\mu_1,0)\otimes (\mu_2,0) -
 (\mu_1+1,0)\otimes (\mu_2-1,0)\right]\eqlabel\pror$$ To change to variables
which carry the Dynkin labels we make the substitution
$M_2 \mapsto M_2M_1^{-1}$, so that $M_1$ now
carries
the first Dynkin label. This introduces negative
powers of $M_1$, corresponding to products (\pror)
 with $\mu_1 < \mu_2$,
which
are not required. So we must keep only non-negative
degree
terms in $M_1$ to obtain the final generating
function.  Denote the resulting expression as $G_4(L_1,L_2,M_1,M_2,N_1,N_2)$; it
reads
$$\eqalign{ 
G_4=~&{(1-L_1L_2M_1M_2N_1N_2)\over
(1-L_1 N_1)(1-L_1M_2)  (1-L_2 M_1) (1-L_2 N_2)}\cr&\times{1\over (1-M_2
N_2) (1-M_1 N_1)(1-L_1 M_1 N_2) (1-L_2 M_2 N_1)} \cr}\eq$$
which is the usual form of the $su(3)$ generating function (cf. section 2.5 of
[\BCM]).

\def\OMun{{\displaystyle\Om_{\geq}^{M_1}}}
\subsec{Extension of the determinantal formula methods to fusion rules: the
$\su(3)$ case}


The starting point for the derivation of 
the  $\su(3)$ fusion generating function
is the generating function for
fusions of the form $$[k-\la_1-\la_2,\lambda_1,\lambda_2]\times
[k-\mu_1,\mu_1,0]\eq$$ These fusions are known explicitly and the
information on their fusion coefficients can be lifted to the following
generating function [\ref{ F. Goodman  and  H. Wenzl, 
Adv. Math. 82 (1990) 244.}\refname\GW]
$$\eqalign{
F_1(&d,L_1,L_2,M_1,N_1,N_2) =\cr
&{1\over(1-d)(1-dL_1N_1)(1-dL_2N_2)(1-dL_2M_1)(1-dM_1N_1)
(1-dL_1M_1N_2)}\cr}
\eq$$

As explained in the previous subsection, the generating function for
products  $$[k-\la_1-\la_2,\lambda_1,\lambda_2]\times
[k-\mu_1-\mu_2,\mu_1,0]\times[k-\mu_2,\mu_2,0]\eq$$ is given by 
$$\eqalign{
F_2(d,L_1,L_2,&M_1,M_2,N_1,N_2)=\cr
&\Oz\,\OR \,F_1(z^{-1}d,L_1,L_2,M_1,R_1^{-1},R_2^{-1})
F_1(z,R_1,R_2,M_2,N_1,N_2).\cr} \eq$$ 
Here the variable $z$ is introduced in order to keep the level fixed in the
composition. 
By the determinantal formula, the 
generating function  is essentially
$$ F_3(d,L_1,L_2,M_1,M_2,N_1,N_2)
=\OMun\,(F_2 - M_2M_1^{-1}F_2)  \eq$$ 
except that the coefficient of $M_1^{\mu_1}M_2^{\mu_2} $
is the multiplicity of $(\mu_1-\mu_2,\mu_2)$.
Thus the final generating function is
$$F_4 = 
\OMun\,  F_3(d,L_1,L_2,M_1,M_2M_1^{-1},N_1,N_2)
\eq$$
This  reproduces the
generating function given in [\CMW] and re-derived
above.

\subsec{Determinantal formula methods applied to the $\sp(4)$ and 
$\su(4)$ cases}

In principle, the above procedure can be used to
calculate the fusion rule generating functions
for
 $\su(4)$ and $\sp(4)$. 
Unfortunately,
the intermediate expressions are too large to be manageable, even when manipulated
with  computer assistance. However, it is possible to calculate
the specialisation of these generating functions with
all but one variable, the level-grading variable, set equal to
1.  For example, in the above calculation for $\su(3)$ we could
have set $L_1=L_2=N_1=N_2=1$ at the start of the calculation
since they are not needed at any intermediate steps. Similarly
we can set $M_2=M_1^{-1}$ at the last step which has the
effect of setting $M_2=1$ in the final generating function.
If we set all variables equal to 1, except the one that keeps track of
the level, then the resulting generating function $G(d)$ counts the
number of independent couplings at each level.
The  $\su(4)$ and $\sp(4)$ specialised generating functions have been calculated
in this way and the results are:
$$
 G^{\su(4)}(d)= {{d^6  + 4 d^5  + 13 d^4  + 16 d^3  + 13 d^2  + 4d +
1}\over {(1 - d)^{12}   (1 - d^2 )}}
\eqlabel\suqua$$
 and 
$$
 G^{\sp(4)}(d)= { {{d}^{4}+2\,{d}^{3}+5\,{d}^{2}+2\,d+1}\over{\left
(1-d\right)^{9}
\left (1+d\right )}}
\eq$$
These expressions agree with the specialisation of the generating
functions found in sections 6 and 7; this thus  provides a very strong
independent verification of these results.  In particular, it corroborates the
closure of our set of fusion elementary couplings.

Although we will not present the details of this derivation, we would like to draw
attention to some technical issues.   There are potentially two problems which
could arise in using the determinantal expansions. The first problem is that the
determinant may contain terms which have level higher than the initial
representation. For example in $\su(3)$ at level 1 the
determinantal expansion of the representation $(0,1)$
is
$$
(0,1) = \det\pmatrix{(1,0) & (0,0) \cr
                      (2,0) & (1,0) }           
\eq$$
The representation $(2,0)$ is integrable only at level 2 and greater.
However it can be shown, using the modification rules of 
[\Ca],
that all such terms in the determinant vanish
identically in the $\sp(2n)$ and $\su(n)$ fusion rings. Thus, when
computing with the determinantal expansions at a given
level, we need only consider terms corresponding to
representations
which exist at that level.

The second  complication which can arise is in a sense the converse
of the first. There are
representations which occur only at levels strictly
greater than $k$, but which have determinantal expansions
which contain products which are defined at level $k$. This does not occur for
the $su(n)$ determinants. However for $sp(4)$ this problem
can happen. The determinant formula for $sp(4)$
is
$$
(\lambda_1,\lambda_2) =
\det \pmatrix{ (\lambda_1+\la_2,0) & (\lambda_2-1,0)\cr
(\lambda_1+\la_2+1,0) + (\lambda_1+\la_2-1,0)& (\lambda_2,0) +
(\lambda_2-2,0)}.
\eq$$
Take for instance the representation $(0,2)$ which does
not exist for level 1.  However the determinant formula yields
$$
(0,2) = (2,0)\otimes(2,0)-(3,0)\otimes(1,0) -
(1,0)\otimes(1,0)+(2,0)\otimes(0,0)\eq$$
The only product which is defined at level 1 is
$ (1,0)\otimes(1,0) = (0,1) $. Thus
the above determinant yields the following {\it modification rule}:
$(0,2) = - (0,1)$ for $\sp(4)$ at level $1$
(see [\Ca] for more details). 
%
%
%
%
Therefore, before
converting the exponent of
$M_1$ into a Dynkin label, we must ensure that it is
 less than or  equal to the exponent of $d$. This can be achieved by
replacing $M_1$ by $M_1y^{-1}$ and $d$ by $dy$ and
then projecting onto non-negative powers of $y$
and finally setting $y=1$.

\subsec{Duality}


As described in [\Ca] and references therein, there is a duality between
fusion rules for $\su(n)$ at level $k$ and fusion
rules for $\su(k)$ at level $n$. This duality is
somewhat involved when using standard Young tableaux.
However, it can be clearly seen using
contravariant tableaux. This duality
can be used to provide a very nice nontrivial check of the
$\su(4)$ generating function.

As discussed above, if all the grading variables in the $\su(4)$ fusion
generating function are set equal to 1, except
for the one associated to the level, we obtain (\suqua).
However in order to use a duality argument to compare this expression
with other generating functions, it needs some modifications. Duality maps Young
tableaux to conjugate Young tableaux. For example $\su(3)$ at level 4
has
$$
\ST{\STrow{\bv\bv\bv\bv}}
\eq$$
as a possible tableau and this maps to
$$
\ST{\STrow{\bv}\STrow{\bv}\STrow{\bv}\STrow{\bv}}
\eq$$
in $\su(4)$ at level 3. In other words we have to include in
the generating functions the terms corresponding to tableaux which have columns
of length $n$ in $\su(n)$.  If ${\cal F}_n(d,L_1,...)$ stands for the original
$\su(n)$ fusion generating function, then the procedure for incorporating
tableaux augmented by columns of length $n$ -- while maintaining the first row
smaller or equal to $k$ --- amounts to calculate
$$
 g_n(d)\equiv{{\partial^2}\over{\partial x\partial y}}\, 
\,
xy\,
{\cal
F}_n(d\,x\,y,x^{-1}L_1,x^{-1}L_2,...,y^{-1}M_1,y^{-1}M_2,...,N_1,N_2,...)\big\vert_{x=y=1}.
\eq$$
The effect of this operation is to multiply 
$$d^kL_1^{\la_1}L_2^{\la_2}\dots M_1^{\mu_1}M_2^{\mu_2}\dots
N_1^{\nu_1}N_2^{\nu_2}\quad {\rm by} \quad (k-\la_1-\la_2\dots
+1)(k-\mu_1-\mu_2...+1)\eq$$ which is the factor needed to add in all the Young
tableaux with all allowed numbers of columns of length $n$. In other words, the
$\su(3)$ tableau
$\ST{\STrow{\bv\bv}\STrow{\bv}}$ at level 5 should appear in following
equivalent forms:
$$\ST{\STrow{\bv\bv}\STrow{\bv}},\qquad
\ST{\STrow{\bv\bv\bv}\STrow{\bv\bv}\STrow{\bv}},\qquad
\ST{\STrow{\bv\bv\bv\bv}\STrow{\bv\bv\bv}\STrow{\bv\bv}},
\qquad \ST{\STrow{\bv\bv\bv\bv\bv}\STrow{\bv\bv\bv\bv}\STrow{\bv\bv\bv}}\eq$$
that is, it should be counted four times.
Doing this and setting all Dynkin-grading variables equal to 1 leads to the
following generating functions:
$$\eqalign{
&g_0={1\over 1 - d}
\quad g_1={ 1 + d\over(1 - d)^3}
\quad g_2={ 1 + 3 d + d^2\over(1 - d)^6}
\quad g_3={d^4  + 6 d^3  + 10 d^2  + 6 d + 1\over (1 - d)^{10}}\cr
&g_4= {d^{10}  +13d^9 +78d^8 +257d^7 +513d^6 +642d^5 +513d^4 +257d^3 + 78d^2
+13d +1\over (1 - d)^{12}   (1 - d^2 )^3}\cr }
\eq$$
The first two functions above correspond to the limiting algebras $\su(0)$ and
$\su(1)$.  For $\su(0)$, there is only the trivial representation and it occurs
at any level.  Therefore, there is a single coupling at every level and there
are no correction factors: 
$g_0(d) = \sum_kd^k$. The function $g_1$ can be  
constructed by duality. We start with the generating function for $\su(k)$
fusions at level 1.  At level 1, we can ignore all relations between the
elementary couplings; moreover, we can keep track only of those elementary
couplings that occur at level 1: these are the various products involving the
fundamental and the scalar representations.  The truncated generating function
then reads 
$${1\over (1-d) \prod_i
[(1-dL_iN_i)(1-dM_iN_i)]\prod_{i,j}(1-dL_iM_jN_{i+j})}\eq$$ where in the last
series of term, the summation is defined modulo $k$ with the understanding that
$N_k=1$. In this function, we replace $d\rw d\, x\,y, \,L_i\rw L_i/x, M_i\rw M_i/y$,
multiply the result by $xy$, differentiate with respect to $x,y,d$ and set
$x=y=L_i=M_i=N_i=1, d=0$ (to keep only the linear term in $d$).  This gives
$(k+1)^2$. Hence we have
$$g_1(d) = \sum_{k=1}^\infty (k+1)^2\, d^k= { 1 + d\over(1 - d)^3}\eq$$

These functions
$g_n(d)$ display very nice properties:

\noindent 1- the factor $(1-d)$ occurs to the power $(n+2)(n+1)/2$ in
the denominator;

\noindent 2- the numerator polynomial $p_n(d)$ satisfies $p_n(1/d)d^{{\rm deg}(p_n)}
= p_n(d)$;

\noindent 3- $p_n(d)$ has positive coefficients;

\noindent 4- the difference between the degree of the numerator
and denominator is $2n$.

The mere fact that $g_4(d)$ shares the generic properties of the previous
$g_n$ functions is  supporting evidence for the correctness of
the
$\su(4)$ generating function.

The Taylor expansions of the $g_n(d)$ functions read:
$$\matrix{
g_0(d)=&1&+&d&  +&d^2  &+&d^3  &+&d^4  &+&d^5  &+\cdots\cr
g_1(d)=&1 &+&4 d &+&9 d^2  &+&16 d^3  &+&25 d^4  &+&36 d^5  &+\cdots\cr
g_2(d)=& 1 &+&9 d &+&40 d^2  &+&125 d^3  &+&315 d^4  &+&686 d^5 &+\cdots\cr
g_3(d)=&1 &+&16 d &+&125 d^2  &+&656 d^3  &+&2646 d^4  &+&8832 d^5  &+\cdots\cr
g_4(d)= &1 &+&25 d &+&315 d^2  &+&2646 d^3  &+&16720 d^4  &+& 85212 d^5 
&+\cdots\cr }\eq$$
from which duality (i.e. horizontal versus vertical) is completely
manifest. (We stress that the `built-in duality' for obtaining
$g_1$ concerns only the second row and the second column.) In
particular the  first 4 terms 1, 25, 315 and 2646 of the
$\su(4)$ function  match the coefficients of the 5-th column.  This again
provides independent evidence for the correctness of the $\su(4)$ fusion
generating function out of which the function $g_4$ has been constructed.  In
particular, this is a decisive test of the necessity of the extra elementary
coupling $\F$ and an evidence for the absence of further additional
elementary couplings. 

{}From the above functions $g_n(d)$ we can construct the sum
$$f(r,d)= g_0(d) + g_1(d)r + g_2(d)r^2 + g_3(d)r^3 + \dots
\eq$$
where $r$ is the grading variable associated to the rank +1 (i.e., its exponent
is the value of $n$ for $\su(n)$). It satisfies $f(r,d) = f(d,r)$ by duality. 
We speculate that other symmetry properties might be
used to provide an explicit formula for $f(r,d)$.

We can illustrate this dual symmetry in a particular example.
Consider the function
${\tilde g}_n(d)$ that counts the number of couplings of the
representation $[k-1,1,0,0,\dots,0]$ with anything in $\su(n)$ at level $k$. 
Since the Young tableau of
$(1,0,0,\dots,0)$ is invariant under a duality transformation exchanging $k$ and
$n$, by summing up the resulting functions multiplied by $r^n$, one should
produce an expression ${\tilde f}(r,d)$ symmetric in the interchange of $r$ and
$d$.   The function
${\tilde g}_n(d)$ is calculated as follows in terms of the original $\su(n)$
fusion generating function ${\cal
F}_n$: 
$$
{\tilde g}_n(d)\equiv {{\partial^2}\over{\partial M_1\partial x}}
\,x\,{\cal
F}_n(dx,x^{-1}L_1,x^{-1}L_2,...,M_1,1,...,1,1...)\vert_{x=y=1,M_1=0,L_1=\dots=1}
\eq$$
As explained above, the differentiation with respect to $x$ is required in
order to take into account all contributing Young diagrams associated to the
first representation ($\la)$.  The second representation being fixed
to be
$(1,0,0,\dots,0)$, does not require an adjusting multiplication factor.  Setting
the variable $M_1=0$, after having differentiated with respect to it, simply
serves to select the  term linear in $M_1$.  Since the representation
$(1,0,0,\dots,0)$ does not exist for $\su(0)$, ${\tilde g}_0(d) =0$. The
function ${\tilde g}_1(d)$ is found by duality as explained previously.   
 The  first few $\su(n)$ functions ${\tilde g}_n(d)$
are found to be:  
$$
\eqalign{
&{\tilde g}_1(d) = {{2 d - d^2}\over {(1 - d)^2}}
\quad\quad {\tilde g}_2(d) = {{3 d - d^2}\over {(1 - d)^3}}\cr
&{\tilde g}_3(d) = {{4 d - d^2}\over {(1 - d)^4}}
\quad\quad {\tilde g}_4(d) = {{5 d - d^2}\over {(1 - d)^5}}\cr
}\eq$$
Fortunately, the general pattern is clear: the expression of ${\tilde
g}_n$ is easily guessed to be:
$${\tilde g}_n(d) = {{(n+1)d-d^2}\over{(1-d)^{n+1}}}\qquad n\geq 1\eq$$ From this
exact form of $g_n(d)$, we can write down readily the exact expression for the
sum
$$
{\tilde f}(r,d) = \sum_{n=1}^\infty {\tilde g}_n(d) r^n = 
{{ d r (  2 -d -r )}\over{  (1-d-r)^2}}
\eq$$
The result is manifestly invariant under the duality transformation that
interchanges
$r$ and
$d$.  

\appendix{B}{Status of previous conjectures}

In this appendix, we would like to clarify the relation between the present work
and our previous ones and state precisely in what sense our previous conjectures
are either embodied in the present reformulation of the problem or have been
proved.

A general approach to the construction of generating functions for fusion rules
was proposed in [\CMW].  It was based on the following two conjectures:

\n 1)  Every coupling is characterised by a threshold level $k_0$. The
multiplicity of a triple product at level $k$ is given by the number of couplings
with threshold levels $\leq k$.
 
\n 2)  There is a choice of forbidden couplings such that the threshold level of
a coupling is  the sum of the threshold levels of its components. 

As already mentioned, it can be shown  [\KMSW] that conjecture 1 is  a
consequence a sharpened formulation of the depth rule [\GW].
This leaves us with a single conjecture which we rename:

 \n {\it Conjecture I: There is a choice of forbidden couplings such that the
threshold level of a coupling is obtained from the sum of the threshold levels of
the elementary couplings that appear in its decomposition.} 

In the formulation of conjecture I, the element of `choice' refers to the
fact that both sides of a tensor-product  relation do not always have the same
threshold level and which one is taken as the forbidden coupling makes a
difference in the generating function for fusion rules. With the notion of a set
of elementary fusion couplings, which includes the scalar one (this is a new
feature of the present work), all  relations acquire equal threshold levels and
this choice becomes immaterial. This suggests the following modification of
conjecture I:

 \n {\it Conjecture I': The
threshold level of a fusion coupling is read off from its decomposition into the
elementary fusion couplings.}

A interesting aspect of this reformulation of the conjecture is that it embodies
an observation that was presented as a conjecture in [\BKMW], namely that the level
is always minimised.  More precisely, in the choice of forbidden couplings, we should
always forbid the one with higher threshold level.  This `minimal level'
prescription is automatically taken into consideration here since the  relations have
identical levels. If one of the  products appears in the  relation with a factor
$\E_0$, it means that the product without this $\E_0$ factor occurs at a lower
level and it is not forbidden. For instance, the relation $\E_1\E_3\E_5=
\E_0\E_7\E_8$ indicates that the coupling $\E_7\E_8$ appears at level 2. In the
tensor-product relation $E_1E_3E_5= E_7E_8$, we  have thus effectively forbid
the higher-level term of the  relation. 

Once
the notion of fusion elementary couplings in terms of which every coupling can be
decomposed (conjecture I') is introduced, this naturally calls for a
reinterpretation in terms of a {\it fusion basis}.  It is indeed plain that our
conjecture (and the mere existence of threshold level)  boils down the
fundamental conjecture  presented in the text, that is, the
existence of a fusion basis.


\vfill\eject
\centerline{\bf REFERENCES}
\vskip 1cm
\immediate\closeout\refs \vskip 0.5cm
  \message{References}\input references
\vfill\eject

\end

Retire:

 {}From the $\su(4)$ fusion elementary couplings and their linear relations, the
fusion generating function is found to be (incomplete):
$$\eqalign{  G&= {\Eb}_0 \prod_{i=1}^{3} {\Ab}_i
 \times [   \prod_{i=1}^{3}  {\Db}_i {\Db'}_i  {\Cb}_1
{\Cb}_2   {\Fb}_1  +\C_3 \prod_{i=1}^{3}   {\Db}_i
 {\Db'}_i 
 {\Cb}_1  {\Cb}_3  {\Fb}\cr 
&+ \C_2 \C_3 \prod_{i=1}^{3}   {\D}_i  {\D'}_i 
 {\Cb}_2  {\Cb}_3  {\Fb}  
 +\B_2 ( {\Eb}_1  {\Bb}_2  {\Bb}_3  {\Cb}_2  {\Cb}_3
 {\Db}_2  {\Db}_3  {\Db'}_2  {\Db'}_3)\cr  
&+\B_1 ( {\Eb}_2  {\Bb}_1  {\Bb}_3  {\Cb}_1  {\Cb}_3
 {\Db}_1
 {\Db}_3  {\Db'}_1  {\Db'}_3 )
+\B_3 ( {\Eb}_3  {\Bb}_2  {\Bb}_1  {\Cb}_2  {\Cb}_1
 {\Db}_2  {\Db}_1  {\Db'}_1  {\Db'}_2) \cr
 & +(\prod_{i=1}^{3}  {\Bb}_i)
(\E_3  {\Db}_1  {\Db}_2  {\Db'}_1  {\Db'}_2
 {\Eb}_3
 {\Cb}_2 
+\E_1  {\Db}_2  {\Db}_3  {\Db'}_2  {\Db'}_3
 {\Eb}_1
 {\Cb}_3 
+\E_2  \Db_1  {\Db}_3  {\Db'}_1  {\Db'}_3  {\Eb}_2
 {\Cb}_1)\cr
& +(\prod_{i=1}^{3}  {\Bb}_i)  (\C_1 \E_3  {\Db}_1  {\Db}_2
 {\Db'}_1  {\Db'}_2  {\Eb}_3  {\Cb}_1
+\C_2 \E_1  {\Db}_2  {\Db}_3  {\Db'}_2  {\Db'}_3
 {\Eb}_1
 {\Cb}_2 \cr
&+\C_3 \E_2  {\Db}_1  {\Db}_3  {\Db'}_1  {\Db'}_3
 {\Eb}_2
 {\Cb}_3) \cr
&+(\prod_{i=1}^{3}  {\Db}_i  {\Db'}_i)  ( {\Bb}_1
 {\Bb}_2  {\Cb}_1  \Cb_2) (1-\B_1 \C_2 \D_3 \D_1^{'}) \cr
&+(\prod_{i=1}^{3}  {\Db}_i  {\Db'}_i)  ( {\Bb}_2
 {\Bb}_3  {\Cb}_2  {\Cb}_3) (1-\B_2 \C_3 \D_1 \D_2^{'}) \cr
&+(\prod_{i=1}^{3}  {\Db}_i  {\Db'}_i)  ( {\Bb}_1
 {\Bb}_3  {\Cb}_1  {\Cb}_3) (1-\E_0 \B_1 \D_1 \D_3 \D_1^{'}
\D_2^{'})
\cr & +(\prod_{i=1}^{3}  {\Bb}_i  {\Db}_i  {\Db'}_i ) 
 {\Cb}_1(1-\E_0 \B_1 \D_1 \D_3 \D_1^{'} \D_2^{'} ) \cr
& +(\prod_{i=1}^{3}  {\Bb}_i  {\Db}_i  {\Db'}_i ) 
 {\Cb}_2(1-\E_0 \B_1 \D_1 \D_3 \D_1^{'} \D_2^{'} ) \cr
& +(\prod_{i=1}^{3}  {\Bb}_i  {\Db}_i  {\Db'}_i ) 
 {\Cb}_3 (1-\E_0\B_1 \D_1 \D_3 \D_1^{'} \D_2^{'} ) ]\cr} \eq$$
Notice that each term of the generating function contains 12 factors.

\n $\bullet$  How can we eliminate factors of $E_0$? new choice of ordering? More:
consider each term of the generating function and decomposes it into the LR basis and
translates the information on the level in data of the LR tableau. Get the
expression of $k$ obtained previously with triangles?